\documentclass[journal]{IEEEtran}

\usepackage{amsmath,amstext,amsfonts,amssymb,amsthm,bbold} 
\usepackage{graphicx,subfigure,color}
\usepackage{graphicx,color,array} 

\newcommand{\bs}{\boldsymbol}
\newcolumntype{P}[1]{>{\centering\arraybackslash}m{#1}}
\newcommand\numberthis{\addtocounter{equation}{1}\tag{\theequation}}

\newtheorem{lemma}{Lemma}

\newtheorem{proposition}{Proposition}
\newtheorem{theorem}{Theorem}

\newtheorem{remark}{Remark}
\newtheorem{assump}{Assumption}
\newtheorem{claim}{Claim}

\begin{document}

\title{Large system analysis of a GLRT for detection with large sensor arrays 
in temporally white noise}

\author
{
	Sonja Hiltunen,
        Philippe~Loubaton,~\IEEEmembership{Fellow,~IEEE}
	, and~Pascal~Chevalier    
	\thanks
	{
		S. Hiltunen is with Thales-Communications-Security, 4 avenue des
		Louvresses 92622 Gennevilliers Cedex (France) and
		Laboratoire d'Informatique Gaspard Monge (CNRS, Universit\'e
		Paris-Est/MLV), 5 Bd. Descartes 77454 Marne-la-Vall\'ee Cedex 2 (France), and
		CNAM, Laboratoire CEDRIC, 292 rue Saint
		Martin, 75141 Paris Cedex 3 (France), 
		sonja.hiltunen@univ-mlv.fr
	}
	\thanks
	{
		P. Loubaton is with Laboratoire d'Informatique Gaspard Monge (CNRS, Universit\'e Paris-Est/MLV), 
		5 Bd. Descartes 77454 Marne-la-Vall\'ee Cedex 2 (France),
		loubaton@univ-mlv.fr
	}
	\thanks
	{
		P. Chevalier is with Thales-Communications-Security, 4 avenue des
		Louvresses 92622 Gennevilliers Cedex (France) and CNAM, Laboratoire CEDRIC,
		292 rue Saint Martin, 75141 Paris Cedex 3 (France),
		pascal.chevalier@thalesgroup.com } }

\maketitle

\begin{abstract}
This paper addresses the behaviour of a classical multi-antenna GLRT 
test that allows to detect the presence of a known signal corrupted 
by a multi-path propagation channel and by an additive temporally white Gaussian 
noise with unknown spatial covariance matrix. The paper is focused 
on the case where the number of sensors $M$ is large, and of the same 
order of magnitude as the sample size $N$, a context which is modeled
by the large system asymptotic regime $M \rightarrow +\infty$, 
$N \rightarrow +\infty$ in such a way that $M/N \rightarrow c$ for 
$c \in (0,+\infty)$. The purpose of this paper is to study the behaviour of a GLRT statistics 
in this regime, and to show that the corresponding theoretical analysis allows to 
accurately predict the performance of the test when $M$ and $N$ are of the same
order of magnitude.   
\end{abstract}
\begin{keywords}
Multichannel detection, asymptotic analysis, GLRT, random
matrix theory
\end{keywords}
\section{Introduction}
\label{sec:intro}

Due to the spectacular development of sensor networks and acquisition devices, 
it has become common to be faced with multivariate signals of high dimension.
Very often, the sample size that can be used in practice in order to perform statistical inference 
cannot be much larger than the signal dimension. In this context, it is well 
established that a number of fundamental existing statistical signal processing 
methods fail. It is therefore of crucial importance to revisit certain classical 
problems in the high-dimensional signals setting. Previous works in this direction 
include e.g. \cite{mestre2008modified} and \cite{p-val-lou-mes-2012} in source localization 
using a subspace method, or
\cite{bianchi-et-al-2011},\cite{krichtman-nadler-2009}, \cite{nada-sil10},\cite{rao-edelman-2008} 
in the context of unsupervised detection. \\

In the present paper, we address the problem of detecting the presence 
of a known signal using a large array of sensors. We assume that the observations 
are corrupted by a temporally white, but spatially correlated (with unknown  
spatial covariance matrix) additive complex Gaussian noise, and study the generalized likelihood 
ratio test (GLRT).  Although our results can be 
used in more general situations, we focus on the detection of a known synchronization 
sequence transmitted by a single transmitter in an unknown multipath propagation
channel.
The behaviour of the GLRT in this context
has been extensively addressed in previous works, but for the low dimensional 
signal case (see e.g.
\cite{astely-jakobsson-1999},\cite{bliss-2010},\cite{dogandzic-nehorai-2002}, \cite{jiang-stoica-li-2004},\cite{book-kay},\cite{viberg-stoice-ottersten-1997},
\cite{zhou-et-al-icc-2012}).
The asymptotic behaviour of the relevant statistics has thus been studied in the
past, but it has been assumed that the number of samples of the training
sequence $N$ converges towards $+\infty$ while the number of sensors $M$ remains fixed.
This is a regime which in practice makes sense when $M << N$. When the number
of sensors $M$ is large, this regime is however often unrealistic, since in
order to avoid wasting resources, the size $N$ of the training 
sequence is usually chosen of the same order of magnitude as $M$. Therefore, we
consider in this paper the asymptotic regime in which both $M$ and $N$ converge towards $\infty$ at the same rate. \\

We consider both the case where the number of paths $L$ remains fixed, 
and the case where $L$ converges towards $\infty$ at the same rate as $M$ and
$N$.
When $L$ is fixed, we prove that the GLRT statistics $\eta_N$ converges under hypothesis 
$\mathrm{H}_0$ towards a Gaussian distribution with mean $ L \log \frac{1}{1- M/N}$ and variance 
$\frac{L}{N} \frac{ M/N}{1 -M/N}$. This is in contrast with the standard asymptotic 
regime $N \rightarrow +\infty$ and $M$ fixed in which the distribution of $\eta_N$ 
converges towards a $\chi^{2}$ distribution. Under hypothesis $\mathrm{H}_1$,  
we prove that $\eta_N$ has a similar behaviour than in the standard asymptotic 
regime $N \rightarrow +\infty$ and $M$ fixed, except that the terms 
$ L \log \frac{1}{1- M/N}$ and $\frac{L}{N} \frac{M/N}{1 - M/N}$ 
are added to the asymptotic mean and the asymptotic variance, respectively. 
When $L$ converges towards $\infty$ at the same rate as $M$ and $N$, we use
existing results (see \cite{bai-silverstein-book} and \cite{zheng-2012})
characterizing the behaviour of linear statistics of the eigenvalues  
of large multivariate $F$--matrices, and infer that the distribution of $\eta_N$ 
under $\mathrm{H}_0$ is also asymptotically Gaussian. The asymptotic mean converges towards 
$\infty$ at the same rate as $L,M,N$ while the asymptotic variance is a
$\mathcal{O}(1)$ term. The asymptotic behaviour of $\eta_N$ under hypothesis $\mathrm{H}_1$ when $L$ 
scales with $M,N$ is not covered by the existing literature. 
The derivation of the corresponding new mathematical results 
would need an extensive work that is not in the scope of the present paper. 
We rather propose a pragmatic approximate distribution for $\eta_N$, 
motivated by the additive structure of its asymptotic mean and variance in the regime 
where $L$ is fixed.  

We evaluate the accuracy of the various Gaussian approximations by numerical
simulations, by comparing the asymptotic means and variances with their
empirical counterparts evaluated by Monte-Carlo simulations. 
Further, we compare the ROC curves corresponding 
to the various approximations with the empirical ones. 
The numerical results show that the standard  approximations 
obtained when $N \rightarrow+\infty$ and $M$ is fixed completely fail if
$\frac{M}{N}$ is greater than $\frac{1}{8}$. The large system 
approximations corresponding to a fixed $L$ and 
$L \rightarrow +\infty$ appear reliable for small values 
of $\frac{M}{N}$, and, of course, for larger values of $\frac{M}{N}$. 
For the values of $L,M,N$ that are considered, the approximations
obtained in the regime $L \rightarrow +\infty$ at the same rate as $M$ and $N$
appear to be the most accurate, and the corresponding ROC-curves
are shown to be good 
approximations of the empirical ones. Therefore, 
the proposed Gaussian approximations allow to reliably predict the performance
of the GLRT when the number of array elements is large. \\

This paper is organized as follows. In section
\ref{sec:presentation}, we provide the signal model under hypotheses $\mathrm{H_0}$ and $\mathrm{H_1}$, recall 
the expression of the statistics $\eta_N$ corresponding 
to the GLRT, and explain that, in order 
to study $\eta_N$, assuming that the additive noise is spatially white and that
the training sequence matrix is orthogonal is not a restriction.
In section \ref{sec:standard}, 
we recall the asymptotic behaviour of $\eta_N$ in the traditional asymptotic regime 
$N \rightarrow +\infty$ and $M$ fixed. 
The main results of this paper, concerning the asymptotic 
behaviour of $\eta_N$ in the regime $M,N$ converge towards $\infty$ at the same rate,
are presented in section \ref{sec:main-results}. 
In this section, we only give outlines of the proofs,
while providing the remaining technical details in 
Appendices. Section \ref{sec:numerical-results} is devoted to the numerical
results, and section \ref{sec:conclusion} concludes the
paper.
\\

{\bf General notations.} 
For a complex matrix ${\bf A}$, we denote by ${\bf A}^T$ and ${\bf A}^*$ its transpose and its conjugate transpose, and by $\mathrm{Tr}({\bf A})$ and $\|{\bf A}\|$ its trace and spectral norm. 
${\bf I}$ will represent the identity matrix and ${\bf e}_n$ will refer to a vector having all its components equal 
to $0$ except the $n$-th which is equal to $1$.

The real normal distribution with mean $m$ and variance $\sigma^2$ is denoted $\mathcal{N}_{\mathbb{R}}(m,\sigma^2)$. A complex random variable $Z = X + i \,  Y$ follows the distribution $\mathcal{N}_{\mathbb{C}}(\alpha+ i \, \beta,\sigma^2)$ if $X$ and $Y$ are independent with respective distributions 
$\mathcal{N}_{\mathbb{R}}(\alpha, \frac{\sigma^2}{2})$ and $\mathcal{N}_{\mathbb{R}}(\beta, \frac{\sigma^2}{2})$. 

For a sequence of random variables $(X_n)_{n \in \mathbb{N}}$ and a random variable $X$, we write
\begin{align}
	X_n \rightarrow X \, {a.s.} \text{ and } X_n \rightarrow_{\mathcal{D}} X
	\notag
\end{align}
when $X_n$ converges almost surely and in distribution,  respectively, to $X$
when $n \rightarrow +\infty$. Finally, if $(a_n)_{n \in \mathbb{N}}$ is a sequence of positive real numbers, $X_n = o_{P}(a_n)$ will stand for the convergence of  $(X_n / a_n)_{n \in \mathbb{N}}$ to $0$ in probability, and  
$X_n = \mathcal{O}_{P}(a_n)$ denotes boundedness in probability (i.e. tightness)
of the sequence $(X_n / a_n)_{n \in \mathbb{N}}$.

\section{Presentation of the problem.}
\label{sec:presentation}

In the following, we assume that a single transmitter sends a 
known synchronization sequence $(s_n)_{n=1, \ldots, N}$ through a fixed channel with
$L$ paths, and that the corresponding signal is received 
on a receiver with $M$ sensors. The received $M$-dimensional signal 
is denoted by $({\bf y}_n)_{n=1, \ldots, N}$. When the transmitter
and the receiver are perfectly synchronized, ${\bf y}_n$ is assumed to 
be given for each $n=1, \ldots, N$ by 
\begin{equation}
\label{eq:signal-model}
{\bf y}_n = \sum_{l=0}^{L-1} {\bf h}_l s_{n-l} + {\bf v}_n
\end{equation}
where $({\bf v}_n)_{n \in \mathbb{Z}}$ is an 
additive independent identically distributed complex Gaussian noise verifying
\begin{align*}
\mathbb{E}({\bf v}_n) =&~ 0 \\
\mathbb{E}({\bf v}_n {\bf v}_n^{T}) =&~ 0 \\
\mathbb{E}({\bf v}_n {\bf v}_n^{*})
=&~ {\bf R} =~ \sigma^{2} \tilde{{\bf R}} \numberthis
\end{align*}
where ${\bf R} > 0$ and $\frac{1}{M} \mathrm{Tr}(\tilde{{\bf R}}) = 1$. Denoting by 
${\bf H}$ the $M \times L$ matrix ${\bf H} = ({\bf h}_0, \ldots, {\bf h}_{L-1})$, 
the received signal matrix ${\bf Y} = ({\bf y}_1, \ldots, {\bf y}_N)$ 
in the presence of a useful signal 
can be written as
\begin{equation} 
\label{eq:matrix-model-H1}
{\bf Y} = {\bf H} {\bf S} + {\bf V}
\end{equation}
where ${\bf V} = ({\bf v}_1, \ldots, {\bf v}_N)$ and where ${\bf S}$ represents
the known signal matrix.  We assume from now on that the size $N$ of the training 
sequence satisfies $N > M+L$.
We remark that the forthcoming results are valid 
as soon as the matrix collecting the observations can be written 
as in Eq. (\ref{eq:matrix-model-H1}). In particular, by appropriately modifying
the matrices ${\bf H}$ and ${\bf S}$, this system model
can equivalently be used for a link with multiple transmit antennas.

Furthermore, in the absence of a useful signal, the
received signal matrix is given by
\begin{equation}
\mathbf{Y} = \mathbf{V}.
\label{eq:matrix-model-H0}
\end{equation}
In this paper, we study the classical problem 
of testing the hypothesis $\mathrm{H}_1$ characterized by Equation
\eqref{eq:matrix-model-H1} against the hypothesis $\mathrm{H}_0$ defined by 
equation \eqref{eq:matrix-model-H0}, in the aim of testing whether
there is a useful signal present in the received signal. The hypotheses are
\begin{align*}
\mathrm{H_0} &: {\bf Y} =  {\bf V} \\
\mathrm{H_1} &: {\bf Y} =  {\bf H}{\bf S} + {\bf V}, \numberthis
\end{align*}
where we assume from now on that ${\bf H}$ and ${\bf R}$ are 
unknown at the receiver side. 
In the following, we will review the expression of the corresponding
generalized maximum likelihood test (GLRT) derived in \cite{bliss-2010}. 
The generalized likelihood ratio $r_N$ is defined by \cite{book-kay}
\begin{equation}	
r_N = \frac{
\underset{\mathbf{R}, \mathbf{H}}{\operatorname{max}}
\, p_{H_1}(\mathbf{Y}~|~\mathbf{S}, \mathbf{H}, \mathbf{R})
} 
{
\underset{\mathbf{R}}{\operatorname{max}}
\, p_{H_0}(\mathbf{Y}~|~\mathbf{R})
}
\label{eq:glrt2-def}.
\end{equation}
The probability density functions are given by
\begin{align*}
p_{H_0}(\mathbf{Y}~|~\mathbf{R})=&\frac{1}{\pi^{NM} \left(\det(\mathbf{R})\right)^{N}}
e^{-\mathrm{Tr}[\mathbf{Y}^* \mathbf{R}^{-1} \mathbf{Y}]} \numberthis \\
p_{H_1}(\mathbf{Y}|~\mathbf{S}, \mathbf{H}, \mathbf{R})=&\frac{1}{\pi^{NM}
\left(\det(\mathbf{R})\right)^{N}}  \cdot \\ 
& \cdot
e^{-\mathrm{Tr}[(\mathbf{Y}-\mathbf{H} \mathbf{S})^* \mathbf{R}^{-1}
(\mathbf{Y}- \mathbf{H} \mathbf{S})]}.
\end{align*}
The first step to calculate $r_N$ is to 
determine $\hat{{\bf R}}_1$ and $\hat{\bf H}$, the ${\bf R}$ and ${\bf H}$ that
maximize the numerator,
and $\hat{{\bf R}}_0$, the ${\bf R}$ that
maximizes the denominator, of equation \eqref{eq:glrt2-def}.
Straightforward calculations show that $\hat{\mathbf{H}} =
\frac{\mathbf{Y}\mathbf{S}^*}{N} (\frac{\mathbf{S}\mathbf{S}^*}{N})^{-1}$ 
and $\hat{\mathbf{R}}_1 = \frac{\mathbf{Y}\mathbf{Y}^*}{N} -
(\frac{\mathbf{Y}\mathbf{S}^*}{N}) (\frac{\mathbf{S}\mathbf{S}^*}{N})^{-1}
(\frac{\mathbf{S}\mathbf{Y}^*}{N})$. 
Similarly, $\hat{{\bf R}}_0$ is given
by $\hat{\mathbf{R}}_0 = \frac{\mathbf{Y}\mathbf{Y}^*}{N}$.

Inserting these estimates into equation \eqref{eq:glrt2-def}
leads to $r_N = \left(\mathrm{det}(\hat{{\bf R}}_1  \hat{{\bf R}}_0^{-1})
\right)^{-N}$.
Therefore, the log-likelihood ratio $\eta_N$, defined by $\eta_N = \frac{\log
r_N}{N}$, is given by
\begin{align}
\eta_N = - \log \mathrm{det} \left[ {\bf I}_M - \hat{{\bf R}}_0^{-1/2}
\frac{{\bf Y}{\bf S}^{*}}{N} \left(\frac{{\bf S}{\bf S}^{*}}{N}\right)^{-1} \frac{{\bf S}{\bf Y}^{*}}{N} \hat{{\bf R}}_0^{-1/2} \right] 
\end{align}
or, using the identity $\mathrm{det}({\bf I} - {\bf A} {\bf B}) = \mathrm{det}({\bf I} - {\bf B} {\bf A})$, by 
\begin{equation}
\label{eq:def-eta}
\eta_N  = -\log \mathrm{det} \left[ {\bf I}_L -  {\bf T}_N \right]
\end{equation}
where ${\bf T}_N$ is the $L \times L$ matrix defined by 
\begin{equation}
\label{eq:def-T} 
{\bf T}_N = \left(\frac{{\bf S}{\bf S}^{*}}{N}\right)^{-1/2}  \frac{{\bf S}{\bf Y}^{*}}{N} \left(\frac{{\bf Y}{\bf Y}^{*}}{N} \right)^{-1} 
\frac{{\bf Y}{\bf S}^{*}}{N} \left(\frac{{\bf S}{\bf S}^{*}}{N}\right)^{-1/2} 
\end{equation}
The generalized maximum likelihood test consists then in comparing 
$\eta_N$ to a threshold.\\

In order to study the behaviour of the test in Eq. (\ref{eq:def-eta}),
we study the limit distribution of $\eta_N$ under each hypothesis. 
For this, we remark that it is possible 
to assume without restriction that $\frac{{\bf S}{\bf S}^{*}}{N} = 
{\bf I}_L$ is verified and that $\mathbb{E}({\bf v}_n {\bf v}_n^{*}) =
\sigma^{2} {\bf I}$, i.e. $\tilde{{\bf R}}$ is reduced to the identity matrix.
If this is not the case, we denote by $\tilde{{\bf S}}$ the matrix 
\begin{align}
\tilde{{\bf S}} = \left(\frac{{\bf S}{\bf S}^{*}}{N}\right)^{-1/2} \, {\bf S}
\end{align}
and by $\tilde{{\bf Y}}$ and $\tilde{{\bf V}}$ the whitened observation and noise matrices 
\begin{align*}
\tilde{{\bf Y}} =& \tilde{{\bf R}}^{-1/2} \, {\bf Y}, \\
\tilde{{\bf V}} =&
\tilde{{\bf R}}^{-1/2} \, {\bf V} \numberthis
\end{align*}
It is clear that $\frac{\tilde{{\bf S}}\tilde{{\bf S}}^{*}}{N} = {\bf I}_L$ and that 
$\mathbb{E}(\tilde{{\bf v}}_n  \tilde{{\bf v}}_n^{*}) = \sigma^{2} {\bf I}$. Moreover,  
under $\mathrm{H_0}$, it holds that $\tilde{{\bf Y}} = \tilde{{\bf V}}$, while under $\mathrm{H_1}$, 
$\tilde{{\bf Y}} = \tilde{{\bf H}} \tilde{{\bf S}} + \tilde{{\bf V}}$
where the channel matrix $\tilde{{\bf H}}$ is defined by 
\begin{align}
\tilde{{\bf H}} =  \tilde{{\bf R}}^{-1/2} \,{\bf H} \, ({\bf S}{\bf S}^{*}/N)^{1/2}
\end{align}
Finally, it holds that the statistics $\eta_N$ can also be written as 
\begin{align}
\eta_N = -\log \mathrm{det} \left[ {\bf I}_L - \frac{\tilde{{\bf S}}\tilde{{\bf Y}}^{*}}{N} \left(\frac{\tilde{{\bf Y}}\tilde{{\bf Y}}^{*}}{N} \right)^{-1} 
\frac{\tilde{{\bf Y}}\tilde{{\bf S}}^{*}}{N} \right]
\end{align}
This shows that it is possible to replace ${\bf S}$, $\tilde{{\bf R}}$ and ${\bf H}$ 
by $\tilde{{\bf S}}$, ${\bf I}$, and $\tilde{{\bf H}}$ without modifying the 
value of statistics $\eta_N$. Therefore, without restriction, we assume from now on that 
\begin{equation}
\label{eq:simpler-hypotheses}
\frac{{\bf S}{\bf S}^{*}}{N} = {\bf I}_L, \; \tilde{{\bf R}} = {\bf I}_M
\end{equation}
In the following, we denote by 
${\bf W}$ a $(N-L) \times N$ matrix for which the 
matrix $\boldsymbol{\Theta} = ({\bf W}^{T}, \frac{{\bf S}^{T}}{\sqrt{N}})^{T}$ is unitary 
and define the $M \times (N-L)$ and $M \times L$ matrices ${\bf V}_1$ and ${\bf
V}_2$ by
\begin{equation}
\label{eq:def-V1-V2}
({\bf V}_1, {\bf V}_2)  = {\bf V} \boldsymbol{\Theta}^{*} = ({\bf V} {\bf W}^{*}, {\bf V} \frac{{\bf S}^{*}}{\sqrt{N}})
\end{equation}
It is clear that ${\bf V}_1$ and ${\bf V}_2$ are complex Gaussian random matrices with independent identically distributed $\mathcal{N}_{\mathbb{C}}(0, \sigma^{2})$ 
entries, and that the entries of ${\bf V}_1$ and ${\bf V}_2$ are mutually independent.
We notice that since $N > M+L$, the matrix $\frac{{\bf V}_1 {\bf V}_1^{*}}{N}$
is invertible almost surely. We now express the statistics $\eta_N$ in terms of ${\bf V}_1$ and ${\bf V}_2$. 
We observe that 
\begin{equation}
\label{eq:expre-VV*} 
\frac{{\bf V} {\bf V}^{*}}{N} = \frac{{\bf V}_1 {\bf V}_1^{*}}{N} + \frac{{\bf V}_2 {\bf V}_2^{*}}{N}
\end{equation}
and that 
\begin{align}
\frac{{\bf V} {\bf S}^{*}}{N} = \frac{1}{\sqrt{N}} \left( {\bf V}_1, {\bf V}_2 \right) 
\left( \begin{array}{c} {\bf W} \\  \frac{{\bf S}}{\sqrt{N}} \end{array} \right) \frac{{\bf S}^{*}}{\sqrt{N}}
\end{align}
coincides with $\frac{{\bf V}_2}{\sqrt{N}}$ because ${\bf W} \frac{{\bf S}^{*}}{\sqrt{N}} = 0$. 
Therefore, under hypothesis $\mathrm{H_0}$, $\eta_N$ can be written as
\begin{align}
\eta_N =  -\log \mathrm{det} \left( {\bf I} - \frac{{\bf V}_2^{*}}{\sqrt{N}}
\left( \frac{{\bf V}_1 {\bf V}_1^{*}}{N} + \frac{{\bf V}_2 {\bf V}_2^{*}}{N}
\right)^{-1} \frac{{\bf V}_2}{\sqrt{N}} \right) 
\vspace{-1cm}
\end{align}
Using the identity 
\begin{multline}
\label{eq:lemme-inversion}
{\bf A}^{*} \left( {\bf B} {\bf B}^{*} + {\bf A} {\bf A}^{*} \right)^{-1} = \\
{\bf A}^{*} ( {\bf B} {\bf B}^{*})^{-1} {\bf A} \left( {\bf I} + 
{\bf A}^{*} ( {\bf B} {\bf B}^{*})^{-1} {\bf A} \right)^{-1}
\end{multline}
we obtain that, under hypothesis $\mathrm{H_0}$, $\eta_N$ can be written as
\begin{equation}
\label{eq:expre-eta-H0}
\eta_N = \log \mathrm{det} \left( {\bf I}_L + {\bf V}_2^{*}/ \sqrt{N} \, \left( {\bf V}_1 {\bf V}_1^{*} / N \right)^{-1} \, {\bf V}_2 / \sqrt{N} \right)
\end{equation}
Similarly, it is easy to check that, under $\mathrm{H_1}$, $\eta_N$ is given by 
\begin{equation}
\label{eq:expre-eta-H1}
\eta_N =  \log \mathrm{det} \left( {\bf I}_L + {\bf G}_N \right)
\end{equation}
where the matrix ${\bf G}_N$ is defined by 
\begin{equation}
\label{eq:def-G}
{\bf G}_N = 
\left({\bf H} + {\bf V}_2/ \sqrt{N} \right)^{*} \, \left( {\bf V}_1 {\bf V}_1^{*} / N \right)^{-1} \, \left( {\bf H} +  {\bf V}_2 / \sqrt{N} \right) 
\end{equation}

\section{Standard asymptotic analysis of $\eta_N$.}
\label{sec:standard}
In order to give a better understanding of the 
similarities and differences with the more complicated case where $M$ and $N$ converge towards 
$+\infty$ at the same rate, we first recall some standard results concerning the
asymptotic distribution of $\eta_N$ under $\mathrm{H}_0$ and $\mathrm{H}_1$ 
when $N \rightarrow +\infty$ but $M$ remains fixed. 
\subsection{Hypothesis $\mathrm{H}_0$.}
A general result concerning the GLRT, known as Wilk's theorem (see e.g.
\cite{book-kay}, \cite{book-young-smith} Chapter 8-5), implies that $N \eta_N$ converges in distribution towards 
a $\chi^{2}$ distribution with $2ML$ degrees of freedom. For the reader's convenience, 
we provide an informal justification of this claim. We use
(\ref{eq:expre-eta-H0}) and remark that when $N \rightarrow +\infty$ and $M$ and $L$ remain fixed, 
the matrices ${\bf V}_1 {\bf V}_1^{*}/N$ 
and $\frac{1}{N} {\bf V}_2^{*} \left( {\bf V}_1 {\bf V}_1^{*}/N \right)^{-1} {\bf V}_2$ converge a.s. 
towards $\sigma^{2} {\bf I}$ and the zero matrix respectively. Moreover, 
\begin{align}
\frac{1}{N} {\bf V}_2^{*} \left( {\bf V}_1 {\bf V}_1^{*}/N \right)^{-1} {\bf V}_2 =  \frac{1}{\sigma^{2}}  {\bf V}_2^{*} {\bf V}_2/N + o_P(\frac{1}{N})
\end{align}
and a standard second order expansion of $\eta_N$ leads to
\begin{align}
\eta_N = \frac{1}{\sigma^{2}} \mathrm{Tr} \left( {\bf V}_2^{*} {\bf V}_2/N \right) + o_P(\frac{1}{N})
\end{align}
This implies immediately that the limit distribution of $N \, \eta_N$ is a chi-squared distribution
with $2ML$ degrees of freedom. Informally, this implies that $\mathbb{E}(\eta_N)
\simeq L \frac{M}{N}$ and $\mathrm{Var}(\eta_N) \simeq \frac{L}{N} \frac{M}{N}$.
\subsection{Hypothesis $\mathrm{H}_1$.}
\label{subsec:standard-H1}
Under hypothesis $\mathrm{H}_1$, $\eta_N$ is given by (\ref{eq:expre-eta-H1}).
When $N \rightarrow +\infty$ and $M$ and $L$ remain fixed, 
the matrix ${\bf V}_1 {\bf V}_1^{*} / N $ converges a.s. towards $\sigma^{2}
{\bf I}$ and it is easily seen that
\begin{align*}
\eta_N = \log \mathrm{det} \Big( {\bf I} +& \frac{ {\bf H} {\bf
H}^{*}}{\sigma^{2}} \Big) + \\ 
\mathrm{Tr}\Big[ \Big( {\bf I} + &\frac{ {\bf H} {\bf H}^{*}}{\sigma^{2}}
\Big)^{-1} \boldsymbol{\Delta}_N \Big] +\mathcal{O}_P(1/N) \numberthis
\end{align*} 
where the matrix $\boldsymbol{\Delta}_N$ is given by
\begin{align}
\boldsymbol{\Delta}_N = {\bf H}^{*} \boldsymbol{\Upsilon}_N {\bf H} + 
\frac{1}{\sigma^{2}} \left( \frac{{\bf V}_2^{*}}{\sqrt{N}} {\bf H}  + 
{\bf H}^{*}  \frac{{\bf V}_2}{\sqrt{N}} \right)
\end{align}
with $\boldsymbol{\Upsilon}_N = \left({\bf V}_1 {\bf V}_1^{} / N \right)^{-1} - {\bf I}/\sigma^{2}$. Standard calculations show that 
\begin{align}
\sqrt{N} \left( \eta_N - \log \mathrm{det}\left( {\bf I} + \frac{{\bf H} {\bf H}^{*}}{\sigma^{2}}\right) \right) \rightarrow \mathcal{N}(0,\kappa_1)
\end{align}
where $\kappa_1$ is given by 
\begin{equation}
\label{eq:expre-kappa1}
\kappa_1 = \mathrm{Tr} \left[ {\bf I} - \left({\bf I} + \frac{{\bf H}^{*}{\bf H}}{\sigma^{2}}\right)^{-2} \right]
\end{equation}
Note that in \cite{book-kay} and \cite{zhou-et-al-icc-2012}, the asymptotic 
distribution of $\eta_N$ is studied under the assumption that the entries of
the matrix ${\bf H}$ are $\mathcal{O}(\frac{1}{\sqrt{N}})$ terms. In that
context, $\eta_N$ behaves as a non-central $\chi^{2}$ distribution.

\section{Main results.}
\label{sec:main-results}
In this section, we present the main results of this paper related to the asymptotic 
behaviour of $\eta_N$ when $M$ and $N$ converge towards $\infty$ at the same rate.
The analysis of $\eta_N$ in the asymptotic regime $M$ and $N$ converge towards $\infty$ at the same rate differs deeply from the standard regime studied in section \ref{sec:standard}. In particular, 
it is no longer true that the empirical covariance matrix ${\bf V}_1 {\bf V}_1^{*}/N$ 
converges in the spectral norm sense towards $\sigma^{2} {\bf I}$. This, of
course, is due to the fact that the number of entries of this $M \times M$ matrix is
of the same order of magnitude than the number of available scalar observations 
(i.e. $M(N-L) = \mathcal{O}(MN)$). We also note that for any deterministic $M \times M$ matrix ${\bf A}$,  the diagonal entries of the $L \times L$ matrix 
$\frac{1}{N} {\bf V}_2^{*} {\bf A} {\bf V}_2$ converge towards $0$ when $N \rightarrow +\infty$ and $M$ remains fixed, while this does not hold when $M$ and $N$ are of the same order of magnitude (see 
Proposition \ref{prop:trace-lemma} in Appendix \ref{appendix:useful-technical-results}). It turns out that the asymptotic regime 
where $M$ and $N$ converge towards $\infty$ at the same rate is more complicated than the conventional regime of section \ref{sec:standard}. 
As the proofs of the following theorems are rather technical, we just provide in
this section the outlines of the approaches that are used to establish them. 
The detailed proofs are given in the Appendix \ref{appendix:proof-of-theorems}. 

\subsection{Asymptotic behaviour of $\eta_N$ when the number of paths $L$ remains fixed when 
$M$ and $N$ increase.}
\label{subsec:Lfixed}
All along this section, we assume that: 
\begin{assump}
\label{assump:asymptotic-regime}
\begin{itemize}
\item 
$M$ and $N$ converge towards $+\infty$ in such a way that $c_N = \frac{M}{N} < 1 - \frac{L}{N}$ 
converges towards $c$, where $0< c < 1$ 
\item 
the number of paths $L$ remains fixed when $M$ and $N$ increase.
\end{itemize}
\end{assump}
In the asymptotic regime defined by Assumption \ref{assump:asymptotic-regime}, $M$ 
can be interpreted as a function $M(N)$ of $N$. Therefore, $M$-dimensional vectors or 
matrices where one of the dimensions is $M$ will be indexed by $N$ in the
following. Moreover, in order to 
simplify the exposition, $N \rightarrow +\infty$ 
should be interpreted in this section as the asymptotic regime
defined by Assumption \ref{assump:asymptotic-regime}. \\

As $M$ is growing, we have to be precise with how the power of the useful signal
component ${\bf H} {\bf S}$ is normalized. In the following, we assume that the norms of vectors $({\bf h}_l)_{l=0, \ldots, L-1}$ 
remain bounded when the number of sensors $M$ increases. This implies that 
the signal to noise ratio at the output of the matched filter ${\bf S}^{*} {\bf H}^{*} {\bf Y}/\sqrt{N}$, i.e.  
 $\mathrm{Tr}\left(({\bf H}^{*} {\bf H})^{2} \right) / \left(\sigma^{2} \mathrm{Tr}({\bf H}^{*} {\bf H}) \right)$,  
 is a $\mathcal{O}(1)$ term in our asymptotic regime. We mention however 
that the received signal to noise ratio $ \mathrm{Tr}({\bf H}^{*}  {\bf H}) / (M \sigma^{2})$ converges towards $0$ 
at rate $\frac{1}{N}$ when $N$ increases. 

\subsubsection{Asymptotic behaviour of $\eta_N$ under hypothesis
$\mathrm{H}_0$} 
Under hypothesis $\mathrm{H_0}$, the following theorem holds. 
\begin{theorem}
\label{theo:eta0}
It holds that 
\begin{equation}
\label{eq:convergence-as-eta0}
\eta_N - L \, \log \left(\frac{1}{1-c_N}\right) \rightarrow 0 \, a.s.
\end{equation}
and that 
\begin{equation}
\label{eq:tlc-HO}
\frac{\sqrt{N}}{\sqrt{\frac{L c_N}{1-c_N}}} \, \left(\eta_N -  L \, \log \left(\frac{1}{1-c_N}\right)\right) \rightarrow_{\mathcal D} \mathcal{N}_{\mathbb{R}}(0,1)
\end{equation}
\end{theorem}
Informally, Theorem \ref{theo:eta0} leads to $\mathbb{E}(\eta_N) \simeq - L
\log(1-c_N)$ and $\mathrm{Var}(\eta_N) \simeq \frac{L}{N} \frac{c_N}{1-c_N}$. We recall that if $M$ is fixed, $N \eta_N$ behaves like a $\chi^{2}$ distribution with $2ML$ degrees of freedom. 
In that context, 
$\mathbb{E}(\eta_N) \simeq L c_N$ and $\mathrm{var}(\eta_N) \simeq \frac{L}{N} c_N$. Therefore, the behaviour of 
$\eta_N$ in the two asymptotic regimes deeply differ. However, if $c_N \rightarrow 0$, $-\log(1-c_N) \simeq c_N$, 
and the asymptotic means and variances
of $\eta_N$ tend to coincide. \\

{\bf Outline of the proof.} We denote by $\mathbf{F}_N$ the $L \times L$ matrix
\begin{equation}
\label{eq:def-F}
\mathbf{F}_N = {\bf V}_2^{*}/ \sqrt{N} \, \left( {\bf V}_1 {\bf V}_1^{*} / N \right)^{-1} \, {\bf V}_2 / \sqrt{N}
\end{equation}
and remark that under $\mathrm{H_0}$, (\ref{eq:expre-eta-H0}) leads to 
\begin{equation}
\label{eq:expre-eta-F}
\eta_N = \log \mathrm{det}\left( {\bf I}_L + \mathbf{F}_N \right)
\end{equation}
\underline{First step: proof of (\ref{eq:convergence-as-eta0})}. As $L$ does not increase with $M$ and $N$, 
it is sufficient to establish that 
\begin{equation}
\label{eq:almost-sure-behaviour-F}
{\bf F}_N - \frac{c_N}{1 - c_N} \, {\bf I}_L \rightarrow 0 \; a.s.
\end{equation}
Our approach is based on the observation that if ${\bf A}_N$ is a $M \times M$ 
deterministic Hermitian matrix verifying  $\sup_N \| {\bf A}_N \| < a < +\infty$, then, 
\begin{align*}
\label{eq:lemme-trace}
\mathbb{E}_{{\bf V}_2} \Big| \left({\bf V}_2^{*}/ \sqrt{N} \, {\bf A}_N {\bf
V}_2 / \sqrt{N} \right)_{k,l} \\
- \frac{\sigma^{2}}{N} \mathrm{Tr}({\bf A}_N) & \,
\delta(k-l) \Big|^{4}  \; \leq \frac{C(a)}{N^{2}} \numberthis
\end{align*}
where $C(a)$ is a constant term depending on $a$, and where $\mathbb{E}_{{\bf V}_2}$ 
represents the mathematical expectation operator w.r.t. ${\bf V}_2$. 
This is a consequence of Proposition \ref{prop:trace-lemma} in the Appendix \ref{appendix:useful-technical-results}. 
Assume for the moment that there exists 
a deterministic constant $a$ such that 
\begin{equation}
\label{eq:condition-fausse}
\| \left(  {\bf V}_1 {\bf V}_1^{*} / N \right)^{-1}  \| \leq a
\end{equation}
for each $N$ greater than a non random integer $N_0$. Then, as 
${\bf V}_1$ and ${\bf V}_2$ 
are independent, it is possible to use (\ref{eq:lemme-trace}) for ${\bf A}_N = \left( {\bf V}_1 {\bf V}_1^{*} / N \right)^{-1}$ and to take the mathematical expectation w.r.t. ${\bf V}_1$ of 
(\ref{eq:lemme-trace}) to obtain that 
\begin{equation}
\label{eq:lemme-trace-nonconditionne}
\mathbb{E} \left|  \left({\bf F}_N \right)_{k,l}  - \frac{\sigma^{2}}{N} \mathrm{Tr}\left(  {\bf V}_1 {\bf V}_1^{*} / N \right)^{-1} \, \delta(k-l) \right|^{4} \leq  \frac{C(a)}{N^{2}}
\end{equation}
for each $N > N_0$, and, using the Borel-Cantelli lemma, that 
\begin{equation}
\label{eq:consequence-lemme-trace-nonconditionne}
{\bf F}_N - \frac{\sigma^{2}}{N} \mathrm{Tr}\left(  {\bf V}_1 {\bf V}_1^{*} / N \right)^{-1} \,  {\bf I}_L \rightarrow 0 \; a.s.
\end{equation}
In order to conclude, we use known results related to the almost sure convergence of the eigenvalue distribution 
of matrix ${\bf V}_1 {\bf V}_1^{*} / N$ towards the so-called Marcenko-Pastur distribution (see Eq. (\ref{eq:comportement-as-trace-Q}) in the Appendix \ref{appendix:useful-technical-results} ) which imply that 
\begin{align} 
\frac{1}{N} \, \mathrm{Tr}\Big(  {\bf V}_1 {\bf V}_1^{*} / N \Big)^{-1} 
-  \frac{c_N}{\sigma^{2}(1 - c_N)} \rightarrow 0
\end{align}
almost surely. This, in conjunction with (\ref{eq:consequence-lemme-trace-nonconditionne}), leads to (\ref{eq:almost-sure-behaviour-F}) 
and eventually to (\ref{eq:convergence-as-eta0}). 

However, there does not exist a deterministic constant $a$ satisfying
(\ref{eq:condition-fausse}) for each $N$ greater than a non random integer. In order to solve this issue, it is sufficient to replace matrix $ \left(  {\bf V}_1 {\bf V}_1^{*} / N \right)^{-1}$ 
by a convenient regularized version. It is well known (see Proposition \ref{prop:convergence-extreme-eigenvalues} in the Appendix \ref{appendix:useful-technical-results}) that the smallest and the largest eigenvalue of
${\bf V}_1 {\bf V}_1^{*} / N$ converge almost surely towards $\sigma^{2}(1 - \sqrt{c})^{2} > 0$ and $\sigma^{2}(1+\sqrt{c})^{2}$ respectively. This implies that if $\mathcal{E}_N$ is the event defined by 
\begin{align*}
\label{eq:definition-EN}
\mathcal{E}_N = \{ & \mbox{one of the eigenvalues of } 
{\bf V}_1 {\bf V}_1^{*} / N  
\mbox{ escapes from } \\
& [\sigma^{2}(1 - \sqrt{c})^{2} - \epsilon, \sigma^{2}(1 + \sqrt{c})^{2} +
\epsilon] \} \numberthis
\end{align*}
(where $\epsilon$ is chosen such that $\sigma^{2}(1 - \sqrt{c})^{2} -
\epsilon > 0$) then, almost surely, for $N$ larger than a random integer, 
it holds that $\mathbb{1}_{\mathcal{E}_N^{c}} = 1$. Therefore, almost surely, 
for $N$ large enough, it holds that $\eta_N = \eta_N \, \mathbb{1}_{\mathcal{E}_N^{c}}$. 
These two random variables thus share the same almost sure asymptotic behaviour.
Moreover, it is clear that $\eta_N \, \mathbb{1}_{\mathcal{E}_N^{c}}$ coincides with $\log \mathrm{det}({\bf I} + {\bf F}_N \mathbb{1}_{\mathcal{E}_N^{c}})$.
In order to study the almost sure behaviour of $\eta_N \mathbb{1}_{\mathcal{E}_N^{c}}$, it is thus sufficient to evaluate 
the behaviour of matrix ${\bf F}_N \mathbb{1}_{\mathcal{E}_N^{c}}$, which has the same 
expression than ${\bf F}_N$, except that matrix  
$ \left(  {\bf V}_1 {\bf V}_1^{*} / N \right)^{-1}$ is replaced by 
$ \left(  {\bf V}_1 {\bf V}_1^{*} / N \right)^{-1} \,  \mathbb{1}_{\mathcal{E}_N^{c}}$. 
The latter matrix verifies 
\begin{equation}
\label{eq:effet-regularisation}
 \left\| \left(  {\bf V}_1 {\bf V}_1^{*} / N \right)^{-1}  \,  \mathbb{1}_{\mathcal{E}_N^{c}} \right\| \leq \frac{1}{\sigma^{2}(1 - \sqrt{c})^{2} - \epsilon}
\end{equation}
for each integer $N$ almost surely. Therefore, the regularized matrix 
$ \left(  {\bf V}_1 {\bf V}_1^{*} / N \right)^{-1} \,  \mathbb{1}_{\mathcal{E}_N^{c}}$
satisfies (\ref{eq:condition-fausse}) almost surely for each integer $N$ for $a = \frac{1}{\sigma^{2}(1 - \sqrt{c})^{2} - \epsilon}$. This immediately leads to the conclusion 
that ${\bf F}_N \,  \mathbb{1}_{\mathcal{E}_N^{c}}$ has the same almost sure behaviour 
than $\frac{c_N}{1 - c_N} {\bf I}_L   \mathbb{1}_{\mathcal{E}_N^{c}}$, or equivalently
than $\frac{c_N}{1 - c_N} {\bf I}_L$. This, in turn, implies (\ref{eq:convergence-as-eta0}). 

\underline{Second step: proof of (\ref{eq:tlc-HO}).}  As $\eta_N = \eta_N \mathbb{1}_{\mathcal{E}_N^{c}}$ almost surely for $N$ large enough,
the asymptotic distributions of $\sqrt{N} [ \eta_N - L \log(\frac{1}{1-c_N}) ]$ and 
$\sqrt{N} [ \eta_N \mathbb{1}_{\mathcal{E}_N^{c}} - L \log(\frac{1}{1-c_N}) ]$ 
coincide. We thus study the latter sequence of random variables because the presence of the regularization 
factor $\mathbb{1}_{\mathcal{E}_N^{c}}$ allows to simplify a lot the derivations. 

A standard second order expansion of 
$\log\mathrm{det}({\bf I} + {\bf F}_N \mathbb{1}_{\mathcal{E}_N^{c}})$ leads to 
\begin{align*}
 \sqrt{N} [ \eta_N \mathbb{1}_{\mathcal{E}_N^{c}} - L \log(\frac{1}{1-c_N}) ] =&
 \\
 (1 - c_N) \sqrt{N} \Big( \mathrm{Tr}({\bf F}_N
 \mathbb{1}_{\mathcal{E}_N^{c}} - &\frac{c_N}{1- c_N} {\bf I}) \Big) + o_P(1)
 \numberthis
\end{align*}
It is thus sufficient to evaluate the asymptotic behaviour of the characteristic 
function $\psi_{N,0}$ of random variable $\beta_{0,N} = (1 - c_N) \sqrt{N} \left( \mathrm{Tr}({\bf F}_N \mathbb{1}_{\mathcal{E}_N^{c}} - \frac{c_N}{1- c_N} {\bf I}) \right)$ 
defined by $\psi_{N,0}(u) = \mathbb{E}(e^{i u \beta_{N,0}})$. For this, 
we first evaluate $\mathbb{E}_{{\bf V}_2}(e^{i u \beta_{N,0}})$, and 
using Proposition \ref{prop:convergence-speed-traceQ} and Proposition \ref{prop:trace-lemma} 
in Appendix \ref{appendix:useful-technical-results}, 
we establish that $\mathbb{E}_{{\bf V}_2}(e^{i u \beta_{N,0}})$ has the same asymptotic 
behaviour as
\begin{align}
\exp \left[-\frac{u^{2}}{2} \, \sigma^{4} \, L \, (1 - c_N)^{2} c_N  \, 
 \frac{1}{M} \mathrm{Tr}\left( \frac{{\bf V}_1 {\bf V}_1^{*}}{N} \right)^{-2} 
 \mathbb{1}_{\mathcal{E}_N^{c}} \right] 
\end{align}
It is known that $\frac{1}{M} \mathrm{Tr}\left( \frac{{\bf V}_1 {\bf V}_1^{*}}{N} \right)^{-2}$ behaves almost surely as $\frac{1}{\sigma^{4}(1 - c_N)^{3}}$
(see Eq. (\ref{eq:comportement-as-trace-Q^2}) in the Appendix \ref{appendix:useful-technical-results}). From this, we obtain immediately that 
\begin{align}
\psi_{N,0}(u) - \exp \left(-\frac{u^{2}}{2} \, \frac{L c_N}{1 - c_N} \right) \; \rightarrow \; 0
\end{align}
for each $u$, which, in turn, establishes (\ref{eq:tlc-HO}).

\subsubsection{Asymptotic behaviour of $\eta_N$ under hypothesis $\mathrm{H_1}$}
The behaviour of $\eta_N$ under hypothesis $\mathrm{H}_1$ is given by the following result. 
\begin{theorem}
\label{theo:eta1}
It holds that 
\begin{equation}
\label{eq:convergence-as-eta1}
\eta_N -  \overline{\eta}_{N,1} \rightarrow 0 \; a.s.
\end{equation}
where $\overline{\eta}_{N,1}$ is defined by 
\begin{equation}
\label{eq:exp-h1}
\overline{\eta}_{N,1} = L \log\frac{1}{1-c_N} + \log \mathrm{det}\left( {\bf I}+{\bf H}^{*}{\bf H}/\sigma^{2} \right) 
\end{equation}
Moreover, 
\begin{equation}
\label{eq:tlc-H1}
\frac{\sqrt{N}}{\left( \frac{L c_N}{1-c_N} + \kappa_1 \right)^{1/2}}  (\eta_N - \overline{\eta}_{N,1}) \rightarrow_{\mathcal{D}} \; \mathcal{N}_{\mathbb{R}}(0,1)
\end{equation}
where $\kappa_1$ is defined by (\ref{eq:expre-kappa1}). 
\end{theorem}
\begin{remark}
\label{re:additive-structure}
Interestingly, it is seen that the asymptotic mean and variance of $\eta_N$ are equal to the sum of 
the asymptotic mean and variance of $\eta_N$ in the standard regime $N \rightarrow +\infty$ and $M$ fixed,
with the extra terms $L \log\left( \frac{1}{1-c_N}\right)$ and $\frac{L c_N}{N (1-c_N)}$, which coincide with the asymptotic mean and variance of $\eta_N$ under $\mathrm{H}_0$. 
\end{remark}
{\bf Outline of the proof.}
We recall that, under $\mathrm{H_1}$, $\eta_N$ is given by (\ref{eq:expre-eta-H1}). 
As in the proof of Theorem \ref{theo:eta0}, it is sufficient to study the regularized statistics 
$\eta_N \mathbb{1}_{\mathcal{E}_N^{c}}$ which is also equal to 
\begin{align}
\eta_N \mathbb{1}_{\mathcal{E}_N^{c}} = \log \mathrm{det} \left( {\bf I}_L +  \mathbb{1}_{\mathcal{E}_N^{c}} \, {\bf G}_N \right)
\end{align}
\underline{First step: proof of (\ref{eq:convergence-as-eta1}).} 
In order to evaluate the almost sure behaviour of $\eta_N \mathbb{1}_{\mathcal{E}_N^{c}}$, we expand  ${\bf G}_N  \mathbb{1}_{\mathcal{E}_N^{c}}$ as
\begin{multline}
\label{eq:expansion-G}
{\bf G}_N  \mathbb{1}_{\mathcal{E}_N^{c}} = {\bf H}^{*} \left( {\bf V}_1 {\bf V}_1^{*} / N \right)^{-1} {\bf H} \; \mathbb{1}_{\mathcal{E}_N^{c}}  + {\bf F}_N  \mathbb{1}_{\mathcal{E}_N^{c}} \, + \\
({\bf V}_2 / \sqrt{N})^{*} \left( {\bf V}_1 {\bf V}_1^{*} / N \right)^{-1} {\bf H} \; \mathbb{1}_{\mathcal{E}_N^{c}} \, + \\
{\bf H}^{*} \left( {\bf V}_1 {\bf V}_1^{*} / N \right)^{-1}  ({\bf V}_2 / \sqrt{N}) \; 
\mathbb{1}_{\mathcal{E}_N^{c}} 
\end{multline}
The first term of the righthandside of (\ref{eq:expansion-G}) is known to 
behave as $\frac{{\bf H}^{*}{\bf H}}{\sigma^{2}(1 - c_N)}$ (see 
(\ref{eq:comportement-as-forme-quadratique-Q}) in the Appendix \ref{appendix:useful-technical-results}) while the independance between ${\bf V}_1$ and ${\bf V}_2$ implies 
that the third and the fourth terms converge almost surely towards the zero
matrix. 
This is because the fourth-order moments w.r.t. ${\bf V}_2$ of their entries
are $\mathcal{O}(\frac{1}{N^{2}})$ terms. \\
\underline{Second step: proof of (\ref{eq:tlc-H1}).} Using a standard second order 
expansion, we obtain immediately that 
\begin{equation}
\sqrt{N} \left( \eta_N \mathbb{1}_{\mathcal{E}_N^{c}} - \overline{\eta}_{N,1} \right)=
\sqrt{N} \, \mathrm{Tr}\left( {\bf D}_N \boldsymbol{\Delta}_N \right) + o_P(1)
\end{equation} 
where $\boldsymbol{\Delta}_N$ and ${\bf D}_N$ are defined by 
\begin{align}
{\bs \Delta}_N = {\bf G}_N \mathbb{1}_{\mathcal{E}_N^{c}} -  \left( \frac{{\bf H}^{*} {\bf H}}{\sigma^{2}(1 - c_N)} + \frac{c_N}{1-c_N} {\bf I} \right) 
\end{align}
and 
\begin{align}
{\bf D}_{N} =  (1 -c_N)  ({\bf I}_L + {\bf H}^{*}{\bf H}/\sigma^{2})^{-1}
\end{align}
In order to establish (\ref{eq:tlc-H1}), it is therefore sufficient to 
evaluate the asymptotic behaviour of the characteristic function $\psi_{N,1}$
of random variable $\beta_{N,1} = \sqrt{N} \, \mathrm{Tr}\left( {\bf D}_N \boldsymbol{\Delta}_N \right)$. 
We define  $\kappa_N$ and $\omega_N$  by
\begin{align}
\kappa_N = \mathrm{Tr}\left( {\bf C}_{N} \left( {\bf V}_1 {\bf V}_1^{*} / N \right)^{-1} \right)
\end{align}
and 
\begin{multline}
\omega_N = \mathrm{Tr} \left[ {\bf D}_{N}  {\bf F}_N
\mathbb{1}_{\mathcal{E}_N^{c}} \right] + \\
\mathrm{Tr} \left[ {\bf D}_{N}
({\bf V}_2 / \sqrt{N})^{*} \left( {\bf V}_1 {\bf V}_1^{*} / N \right)^{-1}
{\bf H} \; \mathbb{1}_{\mathcal{E}_N^{c}} \right] \, + \\
\mathrm{Tr} \left[ {\bf D}_{N} \,  {\bf H}^{*} \left( {\bf V}_1 {\bf V}_1^{*}
/ N \right)^{-1}  ({\bf V}_2 / \sqrt{N}) \;
\mathbb{1}_{\mathcal{E}_N^{c}} \right]
\end{multline}
where ${\bf C}_{N}$ the $M \times M$ matrix given by 
\begin{align}
{\bf C}_{N} =  (1 -c_N)  {\bf H} ({\bf I}_L + {\bf H}^{*}{\bf H}/\sigma^{2})^{-1}  {\bf H}^{*}
\end{align}
Then, $\beta_{N,1}$ can be written as
\begin{eqnarray}
\label{eq:decomposition-beta}
\beta_{N,1} & = & \sqrt{N} \left( \kappa_N - \frac{\mathrm{Tr}({\bf C}_N)}{\sigma^{2}(1 - c_N)} \right) + \\ \nonumber
&   &  \sqrt{N} \left(\omega_N - \frac{c_N}{1-c_N} \, \mathrm{Tr}({\bf D}_N) \right)
\end{eqnarray}
Using the equation above as well as Proposition
\ref{prop:convergence-speed-traceQ} and 
Proposition \ref{prop:trace-lemma} from Appendix
\ref{appendix:useful-technical-results}, we establish that $\mathbb{E}_{{\bf
V}_2}(e^{i u \beta_{N,1}})$ behaves as
\begin{align}
\exp \left( i u \, \sqrt{N} \left( \kappa_N 
- \frac{\mathrm{Tr}({\bf C}_N)}{\sigma^{2}(1 - c_N)} \right) \right) \, 
\exp( - \frac{u^{2}}{2} \zeta)
\end{align}
where $\zeta = \frac{c_N}{(1-c_N)^{3}} \mathrm{Tr}({\bf D}_N^{2}) + 2 \frac{c_N}{(1-c_N)} \mathrm{Tr}({\bf D}_N^{2} {\bf H}^{*} {\bf H})$. In order to obtain the limiting 
behaviour of $\psi_{N,1}(u)$, it is thus sufficient to evaluate the limit of
\begin{align}
\mathbb{E}_{{\bf V}_1} \left[ \exp \left( i u \, \sqrt{N} \left( \kappa_N 
- \frac{\mathrm{Tr}({\bf C}_N)}{\sigma^{2}(1 - c_N)} \right) \right) \right]
\end{align}
This technical point is addressed in Proposition \ref{prop:comportement-formes-quadratiques}
in Appendix \ref{appendix:useful-technical-results}. 

\begin{remark}
It is useful to recall that the expression of the asymptotic mean and variance of $\eta_N$ 
provided in Theorem \ref{theo:eta1} assumes that $\tilde{{\bf R}} = {\bf I}$ and that
$\frac{{\bf S}{\bf S}^{*}}{N} = {\bf I}$. If this is not the case, 
we have to replace ${\bf H}$ by $\tilde{{\bf R}}^{-1/2} {\bf H} \left( {\bf S}{\bf S}^{*}/N \right)^{1/2}$
in Theorem  \ref{theo:eta1}.
\end{remark}
\begin{remark}
We note that Theorem \ref{theo:eta1} allows to quantify the influence of an overdetermination of $L$ 
on the asymptotic distribution of $\eta_N$ under $\mathrm{H}_1$. 
This analysis is interesting from a practical point of view, since it is not
always possible to know the exact number of paths and their delays.
If $L$ is overestimated, i.e.
if the true number of paths is $L_1 < L$, then, matrix ${\bf H}$ can be written
as ${\bf H} = ({\bf H}_1, 0)$.
We also denote by ${\bf S}_1$ and ${\bf S}_2$ the $L_1 \times N$ and $(L - L_1) \times N$ matrices
such that ${\bf S} = \left({\bf S}_1^T, {\bf S}_2^{T}\right)^{T}$. It is easy to check that the second term of $\overline{\eta}_{N,1}$, i.e. 
\begin{align}
\log \mathrm{det} \left( {\bf I}_L + ({\bf S}{\bf S}^{*}/N)^{1/2}  {\bf H}^{*} \tilde{{\bf R}}^{-1} {\bf H} ({\bf S}{\bf S}^{*}/N)^{1/2} \right)
\end{align}
coincides with 
\begin{align}
\log \mathrm{det} \left( {\bf I}_{L_1} + ({\bf S}_1{\bf S}_1^{*}/N)^{1/2}  {\bf H}_1^{*} \tilde{{\bf R}}^{-1} {\bf H}_1 ({\bf S}_1{\bf S}_1^{*}/N)^{1/2} \right)
\end{align}
and is thus not affected by the overdetermination of $L$. Therefore, choosing $L
> L_1$ increases $\overline{\eta}_{N,1}$ by the factor $(L - L_1) \, \log \left( \frac{1}{1-c_N} \right)$. 
As for the asymptotic variance, it is also easy to verify that $\kappa_1$ is not affected 
by the overdetermination of the number of paths, and that the asymptotic
variance is increased by the factor $(L - L_1) \frac{c_N}{1-c_N}$. It is interesting to notice
that the standard asymptotic analysis of subsection \ref{subsec:standard-H1} does not allow
to predict any influence of the overdetermination of $L$ on the asymptotic 
distribution of $\eta_N$. 
\end{remark}

\subsection{Asymptotic behaviour of $\eta_N$ when the number of paths $L$ converges 
towards $\infty$ at the same rate as $M$ and $N$.}
The asymptotic regime considered in section \ref{subsec:Lfixed} is relevant 
when the number of paths $L$ is much smaller than $M$ and $N$. This hypothesis 
may however be restrictive, so that it is of potential interest to study the following 
regime: 
\begin{assump}
\label{assump:asymptotic-regime-L-infty}
$L,M$ and $N$ converge towards $+\infty$ in such a way that $c_N = \frac{M}{N}$ 
and $d_N = \frac{L}{N}$ converge towards $c$ and $d$, where $0< c + d < 1$ 
\end{assump}
As explained below in Paragraph \ref{para:HO-L-infty}, the behaviour of $\eta_N$ under $\mathrm{H}_0$ in this regime is a consequence of existing results. The behaviour of $\eta_N$ under $\mathrm{H}_1$ is however not covered by the existing litterature. The derivation of the corresponding new mathematical results 
needs extensive work that is not in the scope of the present paper. Motivated 
by the additive structure of the asymptotic mean and variance of $\eta_N$ under 
$\mathrm{H}_1$ under assumption \ref{assump:asymptotic-regime}, we propose 
in Paragraph \ref{para:H1-L-infty} a pragmatic
Gaussian approximation of the distribution of $\eta_N$ under $\mathrm{H}_1$

\subsubsection{Asymptotic behaviour of $\eta_N$ under hypothesis $\mathrm{H_0}$}
\label{para:HO-L-infty}
\begin{theorem}
\label{theo:H0-L-infty}
We define $\tilde{\eta}_N$ by
\begin{align*}
\label{eq:asymptotic-mean-alternative-regime}
\tilde{\eta}_N =& - N ( (1-c_N) \log(1-c_N) \\ 
&+ (1-d_N) \log(1-d_N) ) \\
&+ N (1-c_N- d_N) \log(1-c_N-d_N) \numberthis 
\end{align*}
and $\tilde{\delta}_N$ by
\begin{equation}
\label{eq:asymptotic-variance-alternative-regime}
\tilde{\delta}_N = - \log \left( \frac{ 2 \sqrt{a_N^{2} - b_N^{2}}}{a_N + \sqrt{a_N^{2} - b_N^{2}}} \right) 
\end{equation}
where 
\begin{align}
a_N & = \Big( 1 - \frac{c_N}{1-d_N} \Big)^{2} + \frac{d_N}{1-d_N} \Big( 1 +
\frac{c_N(1-c_N)}{d_N(1-d_N)} \Big) \\
b_N & = 2 \frac{d_N}{1-d_N} \, \sqrt{\frac{c_N(1-c_N)}{d_N(1-d_N)}}
\end{align}
Then, it holds that $\mathbb{E}(\eta_N) = \tilde{\eta}_N + \mathcal{O}(\frac{1}{N})$ 
and that 
\begin{equation}
\label{eq:tlc-H0-L-infty}
\frac{1}{\sqrt{\delta_N}} \left( \eta_N - \tilde{\eta}_N \right) \rightarrow_{\mathcal D} \mathcal{N}_{\mathbb{R}}(0,1) 
\end{equation}
\end{theorem}
{\bf Justification.} The eigenvalues of $\mathbf{F}_N$ coincide with the
non-zero eigenvalues of $  ({\bf V}_2  {\bf V}_2^{*}) / N  \, \left( {\bf V}_1
{\bf V}_1^{*} / N \right)^{-1}$. Therefore, $\eta_N$ appears a linear statistics of the eigenvalues of this matrix.
$  ({\bf V}_2  {\bf V}_2^{*}) / N  \,  \left( {\bf V}_1 {\bf V}_1^{*} / N \right)^{-1}$ is 
a multivariate $F$--matrix. The asymptotic behaviour of the empirical 
eigenvalue distribution of this kind of random matrix as well as the 
corresponding central limit theorems are well established 
(see e.g. Theorem 4-10 and Theorem 9-14 in \cite{bai-silverstein-book} 
as well as \cite{zheng-2012}) when the dimensions of $V_1$ and $V_2$ 
converge towards $+\infty$ at the same rate. Theorem \ref{theo:H0-L-infty} 
follows from these results. 

\begin{remark} 
We notice that the results of Theorem \ref{theo:H0-L-infty} differ deeply 
from the results of Theorem \ref{theo:eta0}. We first remark that 
$\tilde{\eta}_N$, and thus $\mathbb{E}(\eta_N)$, converge towards $\infty$ at the same 
rate that $L,M,N$. Moreover, $\eta_N - \mathbb{E}(\eta_N)$ is an
$\mathcal{O}_P(1)$ term under assumption \ref{assump:asymptotic-regime-L-infty}, while it is 
an $\mathcal{O}_P(\frac{1}{\sqrt{N}})$ term when $L$ does not scale with $M,N$.
However, it is possible to informally obtain the
expressions of the asymptotic mean and variance of $\eta_N$ in Theorem \ref{theo:eta0}
from (\ref{eq:asymptotic-mean-alternative-regime}) and
(\ref{eq:asymptotic-variance-alternative-regime}). 
 For this, we remark that a first order expansion w.r.t. $d_N = \frac{L}{N}$ of 
$\tilde{\eta}_N$ and $\tilde{\delta}_N$ leads to 
\begin{align}
\tilde{\eta}_N  = L \, \left( \log(\frac{1}{1-c_N}) + \mathcal{O}(L/N) \right) 
\end{align}
and to 
\begin{align}
\tilde{\delta}_N = \frac{L}{N} \, \frac{c_N}{1-c_N} + \mathcal{O}\left((L/N)^{2}\right)
\end{align}
which, of course, is in accordance with Theorem \ref{theo:eta0}. 
\end{remark}

\subsubsection{Asymptotic behaviour of $\eta_N$ under hypothesis
$\mathrm{H_1}$} 
\label{para:H1-L-infty} 
Under $\mathrm{H_1}$, $\eta_N$ is a linear statistics of the eigenvalues of 
matrix 
\begin{align}
\left( {\bf H} + {\bf V}_2/\sqrt{N} \right) \left( {\bf H} + {\bf V}_2/\sqrt{N} \right)^{*} 
\left( {\bf V}_1 {\bf V}_1^{*}/N \right)^{-1}
\end{align}
To the best of our knowledge, the asymptotic behaviour of the linear
statistics of the eigenvalues of this matrix has not yet been studied in the 
asymptotic regime where $L,M,N$ converge towards $\infty$ at the same rate. 
It is rather easy to evaluate an approximation of the empirical mean 
of $\eta_N$ under $\mathrm{H_1}$ using the results of \cite{dumont-et-al-2010}.
However, to establish the asymptotic gaussianity of $\eta_N$ 
and the expression of the corresponding variance, 
we need to establish a central limit theorem for linear statistics of the 
eigenvalues of non-zero mean large F-matrices.
This needs an important work that is not in the scope of the present paper,
which is why we propose the following pragmatic approximation of the
distribution of $\eta_N$. 
\begin{claim}
\label{claim:pragmatic}
It is relevant to approximate the distribution of $\eta_N$ under $\mathrm{H}_1$ 
by a real Gaussian distribution with mean 
$\tilde{\eta}_N + \log \mathrm{det}\left( {\bf I}+{\bf H}^{*}{\bf H}/\sigma^{2} \right)$ 
and variance $\tilde{\delta}_N + \kappa_{1} / N$. 
\end{claim}
{\bf Justification of Claim \ref{claim:pragmatic}.}
As mentioned in Remark \ref{re:additive-structure}, when $M,N
\rightarrow \infty$ and $L$ is fixed, under $\mathrm{H_1}$, the asymptotic mean 
$\overline{\eta}_{N,1}$ 
is the sum of the asymptotic mean under $\mathrm{H_0}$ given by \eqref{eq:convergence-as-eta0} and the second term
$\log \mathrm{det}\left( {\bf I}+{\bf H}^{*}{\bf H}/\sigma^{2} \right)$. 
Thus, in the regime where $N,M,L \rightarrow \infty$, 
it seems reasonable to approximate the asymptotic mean of $\eta_N$ 
by the sum of $\tilde{\eta}_N$ defined by
\eqref{eq:asymptotic-mean-alternative-regime} with the second term $\log
\mathrm{det}\left( {\bf I}+{\bf H}^{*}{\bf H}/\sigma^{2} \right)$. We can reason similarly with the variance. The asymptotic variance under $\mathrm{H_1}$, \eqref{eq:tlc-H1}, is the sum of the asymptotic variance under $\mathrm{H_0}$, outlined in
 Theorem \ref{theo:eta0}, and the extra term $\frac{\kappa_1}{N}$. 
 Therefore, the asymptotic variance under $\mathrm{H_1}$ in the regime where $N,M,L \rightarrow
 \infty$ can be approximated by the asymptotic variance under $\mathrm{H_0}$ for the same regime, plus the extra
 term $\frac{\kappa_1}{N}$. The results provided by this approximation are evaluated numerically
 in section \ref{sec:numerical-results}. \\

For the reader's convenience, the main results of this paper 
are summarized in Table \ref{table:main-results}, where $\tilde{\delta}_N$
is given by equation
\eqref{eq:asymptotic-variance-alternative-regime}, 
$\kappa_1$ by equation \eqref{eq:expre-kappa1} and $\tilde{\eta}_N$ by
equation \eqref{eq:asymptotic-mean-alternative-regime}.

\begin{table*}
\centering
\begin{tabular}{|P{1.8cm}|P{6cm}|P{9cm}|}
\hline
Assumption on parameters & Distribution under $\mathrm{H}_0$ & Distribution
under $\mathrm{H}_1$ \\
\hline
(a) Classical, $N \rightarrow \infty$ 
& $\eta_N \sim \frac{1}{N}\chi^{2}_{2ML} 
\newline
\Big( \mathbb{E}[\eta_N] = Lc_N,
\mathrm{Var}[\eta_N] = Lc_N \cdot \frac{1}{N} \Big)$
& $\eta_N \sim \mathcal{N}_{\mathbb{R}}\Big( \log\det\Big( {\bf I} + \frac{{\bf
H}{\bf H}^*}{\sigma^2} \Big) , \frac{\kappa_1}{N}\Big)$
\\
\hline
(b) Proposed, $M,N \rightarrow \infty$ 
& $\eta_N \sim \mathcal{N}_{\mathbb{R}}\Big(L \log \frac{1}{1-c_N},
\frac{Lc_N}{1-c_N} \cdot \frac{1}{N}\Big)$
& $\eta_N \sim \mathcal{N}_{\mathbb{R}}\Big( L \log \frac{1}{1-c_N} + 
\log\det\Big( {\bf I} + \frac{{\bf H}{\bf
H}^*}{\sigma^2} \Big),
\frac{\kappa_1}{N} + \frac{Lc_N}{1-c_N} \cdot \frac{1}{N}\Big)$
\\
\hline
(c) Proposed, $L,M,N \rightarrow \infty$ 
& $\eta_N \sim \mathcal{N}_{\mathbb{R}} \Big( \tilde{\eta}_N , 
\tilde{\delta}_N\Big)$ 
& $\eta_N \sim \mathcal{N}_{\mathbb{R}} \Big( \tilde{\eta}_N + \log\det\Big(
{\bf I} + \frac{{\bf H}{\bf H}^*}{\sigma^2} \Big),
\frac{\kappa_1}{N} + \tilde{\delta}_N \Big)$ 
\\
\hline
\end{tabular}
\caption{Asymptotic distribution of $\eta_N$ for different
assumptions, under $\mathrm{H_0}$
and $\mathrm{H_1}$}
\vspace{-0.8cm}
\label{table:main-results}
\end{table*}

\section{Numerical results.}
\label{sec:numerical-results}
In this section, we validate the relevance of the Gaussian approximations of
section \ref{sec:main-results}. In our numerical experiments, we have calculated
the asymptotic expected values and variances as well as their empirical counterparts, 
evaluated by Monte Carlo simulations with $100.000$ trials. 
In this section, to refer to the different approximations,
we use the (a), (b) and (c) defined in table
\ref{table:main-results}.

The fixed channel ${\bf H}$ is equal to 
${\bf H} = \frac{1}{\left(\mathrm{Tr}({\overline{\bf H}} {\overline{\bf H}}^{*})\right)^{1/2}} \, {\overline{\bf H}}$ 
where ${\overline{\bf H}}$ is a realization of a $M \times L$ 
Gaussian random matrix with i.i.d. $\mathcal{N}_c(0, \frac{1}{M})$ entries. We remark that
$\mathrm{Tr}({\bf H} {\bf H}^{*}) = 1$. 

The rows of the training sequence matrix ${\bf S}$ are chosen as cyclic
shifts of a Zadoff-Chu sequence of length $N$ \cite{chu-1972}. Due to the
autocorrelation properties of Zadoff-Chu sequences, designed so that the correlation between
any shift of the sequence with itself is zero, we have ${\bf S}
{\bf S}^* / N = {\bf I}_L$ if $L \leq N$. 

 \subsection{Influence of $c_N = \frac{M}{N}$ on the asymptotic means and variances.}
 We first evaluate the behaviour of the means and variances 
of the three Gaussian approximations in terms of $c_N =  \frac{M}{N}$. We only
show the results for the asymptotic variance under $\mathrm{H_1}$, but note that the results are
similar for the expected values and under hypothesis $\mathrm{H_0}$.
Figure 1 compares the theoretical variances with
 the empirical variances obtained by simulation, under hypothesis $\mathrm{H}_1$, as a function of $c_N$, the
 ratio between $M$ and $N$. In this simulation, $M=10$, $L=5$ and
 $N=20,40,60,80,160,320$. When $c_N$ is small, the three approximations 
 (a), (b) and (c) give the
 same variance, as expected, and are very close to the empirical variance. 
 When $c_N \geq \frac{1}{8}$, the assumption that $M$ is small compared 
 to $N$ is no longer valid, and the classical asymptotic analysis (a) fails. 
The two large system approximations (b) and (c) provide similar results when
$c_N \leq \frac{1}{4}$, i.e.
when $N = 40$, or equivalently when $\frac{L}{N} \leq \frac{1}{8}$. However, when $N = 20$, 
i.e. $\frac{L}{N} = \frac{1}{4}$, (c), the approximation corresponding to the
regime where $L,M,N$ converge towards $\infty$ leads to a much more accurate prediction
of the empirical variance. We remark that the approximation (c) is also reliable
for rather small values of $L,M,N$, i.e. $L=5, M=10, N=20$. 
We also remark that the regimes (b) and (c) where $M,N$ are of the same order of
magnitude capture the actual performance even when $c_N$ is
small, which, by extension, implies that the standard asymptotic analysis (a)
always performs worse compared to the two large system approximations.
 If $N,M$ increase while $c_N$ stays the
 same, the results will be even closer to the theoretical values, 
 since the number of samples is larger. \\
 
 In the simulations that follow, we will use $c_N=1/2$ with $N=300$,
 $M=150$ and $L=10$, if not otherwise stated.
\begin{figure}[htb]
\label{fig:var-h1-growing-cN}
  \centering
    \centerline{\includegraphics[width=8.5cm]{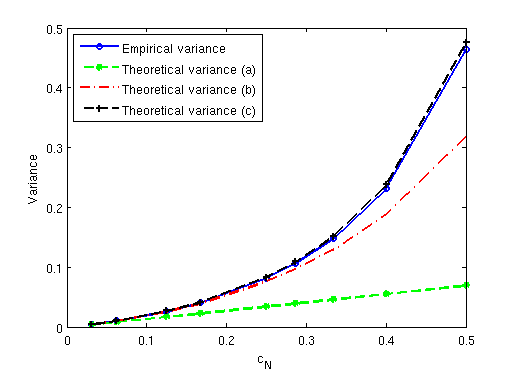}}
\caption{Proposed asymptotic analysis with standard asymptotic analysis}
\end{figure}

\subsection{Comparison of the asymptotic means and variances of the approximations of $\eta_N$ under $\mathrm{H_0}$}

We first compare in figures 
2 and 3
the  asymptotic expected values and variances with the empirical ones when $L$ increases from $L=1$ to $L=30$ while $M=150$ and $N = 300$, 
i.e. $c_N = 1/2$.  
The figures show that the standard asymptotic analysis of section
\ref{sec:standard} completely fails for all values of $L$. This is expected, given
the value of $\frac{M}{N}$. 
As $L$ increases, the assumption that $L$ is small becomes 
increasingly invalid, and the only model that functions well in this 
regime is the model (c). This
is valid both for the expected value and variance, and the
theoretical values are very close to their empirical counterparts.
We remark that the approximation (c), valid 
when $L \rightarrow +\infty$, also allows to capture the actual empirical
performance when $L$ is small.

\begin{figure}[htb]
\label{fig:exp_h0}
\begin{minipage}[b]{1.0\linewidth} 
  \centering
    \centerline{\includegraphics[width=8.5cm]{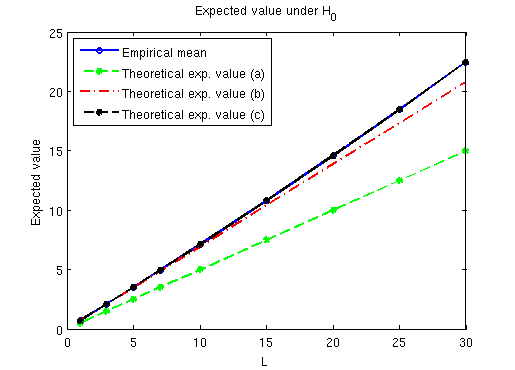}}
\end{minipage}%
\caption{$\mathrm{H_0}$: Asymptotic expected values as a function of L}
\end{figure}
\begin{figure}[htb]
\label{fig:var_h0}
\begin{minipage}[b]{1.0\linewidth}
  \centering
   \centerline{\includegraphics[width=8.5cm]{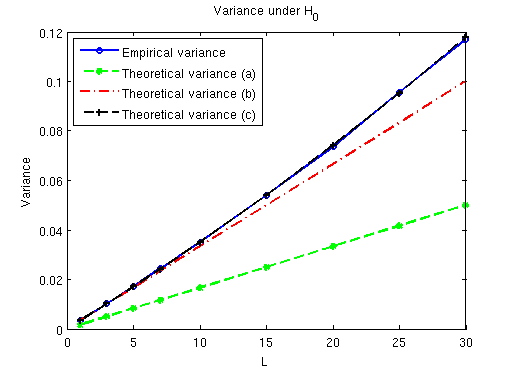}}
\end{minipage}%
\caption{$\mathrm{H_0}$: Asymptotic variances as a function of L}
\end{figure}

\subsection{Validation of asymptotic distribution under $\mathrm{H_0}$}

Although the expected values and variances can be very accurate, this does not
necessarily mean that the empirical distribution is Gaussian. Therefore, we
need to validate also the distribution under $\mathrm{H_0}$. The asymptotic distribution under
$\mathrm{H_0}$ can be validated by analyzing its accuracy when calculating a threshold 
used to obtain ROC-curves. Note that this analysis also shows the applicability 
of the results for a practical case of timing synchronization. 

We calculate the ROC curves in two different ways.
The first is the ROC curve calculated empirically.
We determine a threshold $s$ from the empirical distribution
under $\mathrm{H_0}$ which gives a given probability of false alarm as 
$P_{fa} = \mathbb{P}(\eta_N > s)$. Its corresponding probability
of non-detection, $P_{nd}$, is then
obtained as the probability that the empirical values of the
synchronization statistics under $\mathrm{H_1}$ pass this threshold.
The other ROC-curves are obtained by calculating
the threshold $s$ from the
asymptotic Gaussian distributions under $\mathrm{H_0}$, and using this theoretical threshold to
calculate the $P_{nd}$ from the empirical distribution under $\mathrm{H_1}$. 

Figure 
4 shows the ROC-curves obtained with the
approaches mentioned above when $L=10, M=150, N=300$. 
Since the standard asymptotic analysis
(a) gives very bad results, its results are omitted.  
It is clear that ROC-curve obtained by using the asymptotic distribution (b),
obtained with the assumption that $L$ is small, differs greatly from the results
from the approximation (c), even
for this relatively small value of $L$. This is because 
the theoretical threshold depends greatly on
the expected value, and if it is not precisely evaluated, it gives 
erroneous results. In (c), the model where $N,M,L \rightarrow \infty$, the
expected value and variance are very close to their empirical counterparts, and the
resulting threshold can be used to precisely predict the synchronization 
performance for the set of parameters used when $P_{fa} \geq 10^{-3}$ and $P_{nd} > 10^{-3}$.  
\begin{figure}[htb]
\label{fig:roc-curves}
  \centering
    \centerline{\includegraphics[width=8.5cm]{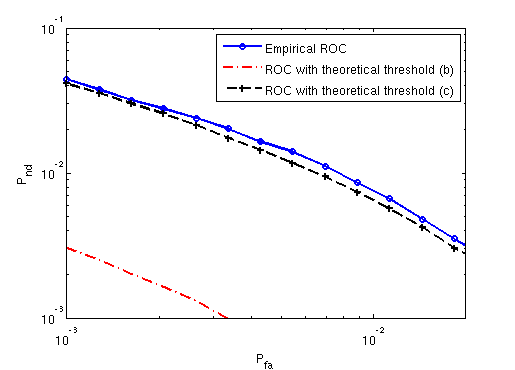}}
\caption{ROC curve obtained with theoretical threshold plotted with the
empirical ROC curve}
\end{figure} 
Figure 
5 shows,
for the regime (c), the ROC curves obtained
with the theoretical threshold, together with the empirical 
results. In the
figure, $L$ goes from 1 to 20, while $M=15L$ goes from 15 to 300 and $N=30L$
goes from 30 to 600. It is seen that when the three parameters grow, 
the distance between the theoretical and empirical ROC curves decreases. 
 \begin{figure}[htb]
\label{fig:roc-theory-threshold-L-M-N-growing}
  \centering
    \centerline{\includegraphics[width=8.5cm]{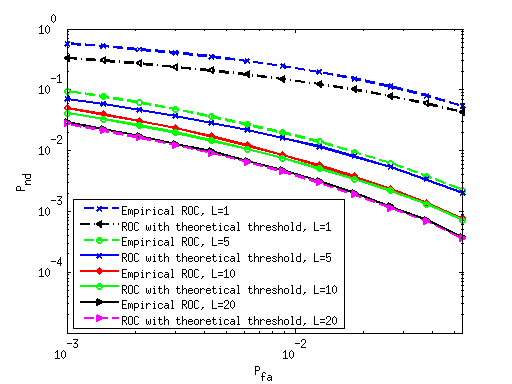}}
\caption{ROC curve obtained with theoretical threshold, for proportionally
growing N,M,L, model (c)}
\end{figure}

\subsection{Comparison of the asymptotic means and variances of the approximations of $\eta_N$ under $\mathrm{H_1}$.}
In this section, we will proceed to validate the expected value and 
variance under $\mathrm{H_1}$. 

Figures 
6 and 7 validate the 
asymptotic expected values and variances under $\mathrm{H_1}$. Similarly to hypothesis
$\mathrm{H_0}$, the theoretical expected values and variances are poorly evaluated using the
standard asymptotic analysis (a). We note that the asymptotic expected values 
deduced for the regime (c)  
are very close to the empirical expected values
and variances. For an $L$ sufficiently small, however, 
also the regime (b) gives asymptotic expected
values and variances that are close to their empirical counterparts. 
\begin{figure}[htb]
  \label{fig:exp_h1}
  \centering
    \centerline{\includegraphics[width=8.5cm]{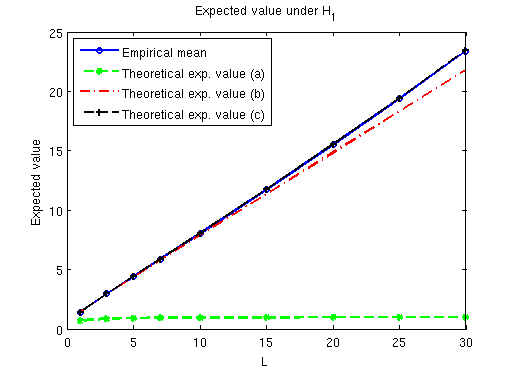}}
\caption{$\mathrm{H_1}$: Asymptotic expected values as a function of L}
\end{figure}
\begin{figure}[htb]
  \label{fig:var_h1} 
  \centering
   \centerline{\includegraphics[width=8.5cm]{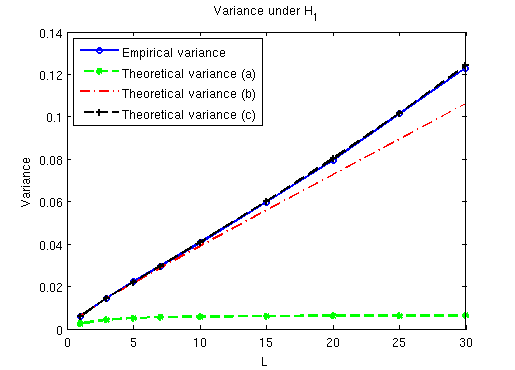}}%
\caption{$\mathrm{H_1}$: Asymptotic variances as a function of L}
\end{figure}

\subsection{Validation of asymptotic distribution under $\mathrm{H_1}$}

To validate the asymptotic distributions under $\mathrm{H_1}$, we calculate theoretical
ROC-curves using both asymptotic distributions. 
For each $P_{fa}$, a threshold $s$ is calculated from the theoretical
Gaussian distribution under $\mathrm{H_0}$. This threshold is then used to calculate the $P_{nd}$ from the
theoretical Gaussian distribution under $\mathrm{H_1}$, using $P_{nd} = 1 -
\mathbb{P}_{H_1}(\eta_{N} > s)$.
Figure 
8 shows these theoretical ROC curves plotted together with the empirical ROC curve. 
Here, $L=10, M=150$ and $N=300$. It is seen that the approximation corresponding to the regime $N,M,L \rightarrow \infty$ provides, as in the context of hypothesis $\mathrm{H_0}$, a more accurate theoretical ROC curve. It is seen 
that the ROC curve associated with the regime small $L$ (b) is closer from the
empirical ROC curve than in the context of hypothesis $\mathrm{H_0}$. This is because the corresponding asymptotic means are, 
for both $\mathrm{H_0}$ and $\mathrm{H_1}$, less than the actual empirical means. These two errors
tend to compensate in the theoretical ROC curves (b), which explains why the
theoretical ROC curve (b) of figure 
8 is more accurate than the corresponding ROC curve of figure 4, for small $L$.

 \begin{figure}[htb]
\label{fig:theoretical-roc}
\begin{minipage}[b]{1.0\linewidth}
  \centering
    \centerline{\includegraphics[width=8.5cm]{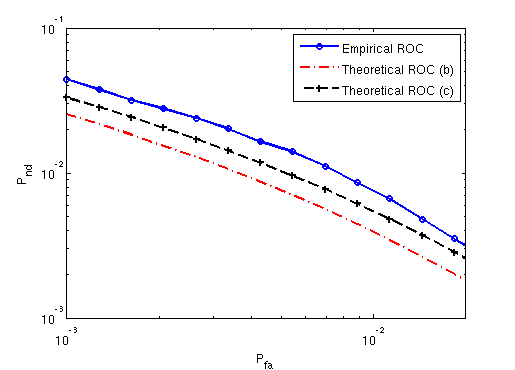}}
\end{minipage}%
\vspace{-0.25cm}
\caption{Theoretical ROC curves plotted with the empirical ROC curve}
\end{figure}
We now evaluate the behaviour of the ROC curves when $N,M,L$ grow at the same rate.
In figure 
9, $L$ goes from 1 to 20, while $M=15L$ goes from 15 to 300 and $N=30L$
goes from 30 to 600. The results show that as $N,M,L$ grow
proportionally, the theoretical results tend to approach the empirical values, 
but that, in contrast with the context of figure
5, a residual error remains. It would be interesting to evaluate more accurately
the asymptotic behaviour of $\eta_N$ under $\mathrm{H_1}$ in 
the regime $L \rightarrow +\infty$, and to check if the 
residual error tends to diminish. However, 
as mentioned in 
Paragraph \ref{para:H1-L-infty}, 
this needs to establish a central limit theorem for linear statistics of the 
eigenvalues of non zero mean large F-matrices, which is a non trivial task. 
\begin{figure}[htb]
\label{fig:theoretical-roc-L-M-N-growing}
\begin{minipage}[b]{1.0\linewidth}
  \centering 
    \centerline{\includegraphics[width=8.5cm]{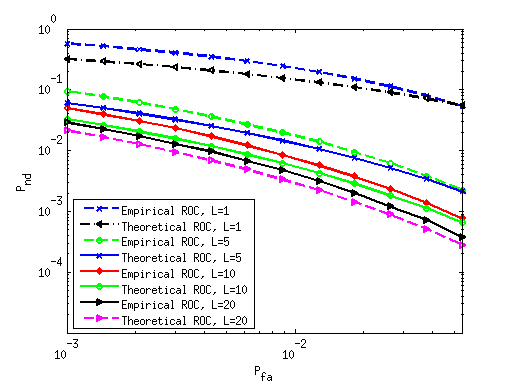}}
    \end{minipage}%
\vspace{-0.25cm}
\caption{Theoretical ROC curves for proportionally growing N,M,L, model (c)}
\end{figure}


\section{Conclusion.}
\label{sec:conclusion}
In this paper, we have studied the behaviour of the multi-antenna GLR detection 
test statistics $\eta_N$ of a known signal corrupted by a multi-path deterministic channel and an additive white Gaussian noise with unknown spatial covariance. We have addressed the case 
where the number of sensors $M$ and the number of samples $N$ of the training sequence 
converge towards $\infty$ at the same rate. When the number of paths $L$ 
does not scale with $M$ and $N$, we have established  that $\eta_N$ has a Gaussian behaviour 
with asymptotic mean $ L \log \frac{1}{1- M/N}$ and variance $\frac{L}{N} \frac{M/N}{1 - M/N}$. 
This is in contrast with the standard regime $N \rightarrow +\infty$ and $M$ fixed 
where $\eta_N$ has a $\chi^{2}$ behaviour. 
Under hypothesis $\mathrm{H_1}$, $\eta_N$ has still 
a Gaussian behaviour. The corresponding asymptotic mean and variance are obtained 
as the sum of the asymptotic mean and variance in the standard regime $N \rightarrow +\infty$ and 
$M$ fixed, and $ L \log \frac{1}{1- M/N}$ and $\frac{L}{N} \frac{M/N}{1 - M/N}$
respectively, i.e. the asymptotic mean and variance under $\mathrm{H_0}$. 
We have also considered the case where the number of paths $L$ converges towards 
$\infty$ at the same rate as $M$ and $N$. Using known results of
\cite{bai-silverstein-book} and \cite{zheng-2012}, concerning the behaviour of linear statistics of the eigenvalues of large F-matrices, we have deduced that in the regime where $L,M,N$ converge to $\infty$ 
at the same rate, $\eta_N$ still has a Gaussian behaviour under $\mathrm{H_0}$, but with a
different mean and variance. The analysis of $\eta_N$ under $\mathrm{H_1}$ when 
 $L,M,N$ converge to $\infty$ needs to establish a central limit theorem for 
linear statistics of the eigenvalues of large non zero-mean F-matrices, a difficult
task that we will address in a future work. Motivated by the results obtained 
in the case where $L$ remains finite, we have proposed to approximate 
the asymptotic distribution of $\eta_N$ by a Gaussian distribution whose mean 
and variance are the sum 
of the  asymptotic mean and variance under $\mathrm{H_0}$ when $L \rightarrow +\infty$ with the asymptotic mean and variance under $\mathrm{H_1}$ in the standard regime $N \rightarrow +\infty$ and $M$ fixed. 
Numerical experiments have shown that the Gaussian approximation corresponding 
to the standard regime $N \rightarrow +\infty$ and $M$ fixed completely fails 
as soon as $\frac{M}{N}$ is not small enough. The large system approximations 
provide better results when $\frac{M}{N}$ increases, while also allowing to
capture the actual performance for small values of $\frac{M}{N}$. We have also observed that, 
for finite values of $L,M,N$, the Gaussian approximation obtained in the regime $L,M,N$ converge towards $\infty$ is more accurate than the approximation in which $L$ is fixed. In particular, 
the ROC curves that are obtained using the former large system approximation 
are accurate approximations of the empirical ones in a reasonable range
of $P_{fa}, P_{nd}$. We therefore believe that our results can be used to
reliably predict the performance of the GLRT, and that the tools that are
developed in this paper are useful in the context of large antenna arrays.

\begin{appendices}

\section{Useful technical results.}
\label{appendix:useful-technical-results}
In this appendix, we provide some useful technical results concerning the
behaviour of certain large random matrices. In the remainder of this 
appendix, ${\bs \Sigma}_N$ represents
a  $M \times N$ matrix with $\mathcal{N}_{\mathbb{C}}(0, \frac{\sigma^{2}}{N})$ i.i.d. elements.
We of course assume in this section that $M$ and $N$ both converge towards $+\infty$ in 
such a way that $c_N = \frac{M}{N} < 1$ converges towards $c < 1$. 
 In the following, we give some results 
concerning the behaviour of the eigenvalues $\hat{\lambda}_{1,N} \leq \hat{\lambda}_{2,N} \ldots \leq \hat{\lambda}_{M,N}$ 
of the matrix  ${\bs \Sigma}_N {\bs \Sigma}_N^{*}$ as well as on its resolvent
${\bf Q}_N(z)$ defined for $z \in \mathbb{C} - \mathbb{R}^{+}$ by 
\begin{equation}
\label{eq:def-resolvent}
{\bf Q}_N(z) = \left( {\bs \Sigma}_N {\bs \Sigma}_N^{*} - z {\bf I}_M \right)^{-1}
\end{equation}
We first state the following classical result (see e.g. \cite{bai-silverstein-book}, Theorem 5.11).
\begin{proposition}
\label{prop:convergence-extreme-eigenvalues}
When $N \rightarrow +\infty$, $\hat{\lambda}_{1,N}$ converges almost surely towards $\sigma^{2}(1 - \sqrt{c})^{2}$
while $\hat{\lambda}_{M,N}$ converges a.s. to $\sigma^{2}(1 + \sqrt{c})^{2}$.
\end{proposition}
In the following, we denote by $\mathcal{I}_{\epsilon}$ the interval defined by 
\begin{equation}
\label{eq:def-Iepsilon}
\mathcal{I}_{\epsilon} = [\sigma^{2}(1 - \sqrt{c})^{2} - \epsilon, \sigma^{2}(1 + \sqrt{c})^{2} + \epsilon]
\end{equation}
(with $\epsilon$ chosen in such a way that $\sigma^{2}(1 - \sqrt{c})^{2} - \epsilon > 0$) and by $\mathcal{E}_N$ the event defined by
\begin{equation}
\label{eq:def-E}
\mathcal{E}_N = \{ \mbox{one of the $(\hat{\lambda}_{k,N})_{k=1, \ldots, M}$ escapes from 
$\mathcal{I}_{\epsilon}$} \}
\end{equation}
and remark that the almost sure convergence of $\hat{\lambda}_{1,N}$ and
$\hat{\lambda}_{M,N}$ implies that
\begin{align*}
\label{eq:evaluation-indicatrice-Ec}
\mathbb{1}_{\mathcal{E}_N^{c}} = 1 \; 
&\mbox{almost surely for each $N$} \\
&\mbox{larger than a random integer} \numberthis
\end{align*}
Proposition \ref{prop:convergence-extreme-eigenvalues} implies that the resolvent ${\bf Q}_N(z)$ 
is almost surely defined on $\mathbb{C} - \mathcal{I}_{\epsilon}$ for $N$ 
large enough, and in particular for $z = 0$. 

Another important property is the almost sure convergence of the empirical 
eigenvalue distribution $\hat{\mu}_N = \frac{1}{M} \sum_{k=1}^M \delta_{\hat{\lambda}_{k,N}}$ of ${\bs \Sigma}_N {\bs \Sigma}_N^{*}$ towards the Marcenko-Pastur distribution
(see e.g. \cite{bai-silverstein-book} and \cite{pastur-shcherbina-book} and the references therein). 
Formally, this means that the Stieltjes transform $\hat{m}_N(z)$ of $\hat{\mu}_N$ defined 
by 
\begin{align}
	\hat{m}_N(z) = \int_{\mathbb{R}} \frac{ d\hat{\mu}_N(\lambda)}{\lambda - z} = \frac{1}{M} \mathrm{Tr}\left({\bf Q}_N(z)\right)
\end{align}
satisfies 
\begin{equation}
\label{eq:convergence-MP}
\lim_{N \rightarrow +\infty} \left( \hat{m}_N(z) - m_{c_N}(z) \right) = 0
\end{equation}
almost surely for each  $z \in \mathbb{C} - \mathbb{R}^{+}$ (and uniformly on each compact subset of $ \mathbb{C} - \mathbb{R}^{+}$), where $m_{c_N}(z)$ represents the Stieltjes transform 
of the Marcenko-Pastur distribution of parameter $c_N$, denoted by $\mu_{c_N}$ in the following.  $m_{c_N}(z)$ 
satisfies the following fundamental equation 
\begin{equation}
\label{eq:stieltjes-MP}
	m_{c_N}(z) = \frac{1}{-z\left(1+\sigma^2 c_N m_{c_N}(z)\right) + \sigma^2 (1-c_N)}
\end{equation}
for each $z \in \mathbb{C}$. $\mu_{c_N}$ is known to be absolutely continuous, its support is the interval 
$[\sigma^2 (1-\sqrt{c_N})^2, \sigma^2 (1+\sqrt{c_N})^2]$, and its density is given by
\begin{align}
	 \frac{\sqrt{\left(x - x_{c_N}^-\right)\left(x_{c_N}^+ - x\right)}}{2 \sigma^2 c_N \pi x} \mathbb{1}_{[x_{c_N}^-,x_{c_N}^+]}(x).
\end{align}
with $x_{c_N}^- = \sigma^2 (1-\sqrt{c_N})^2$ and $x_{c_N}^+ = \sigma^2 (1+\sqrt{c_N})^2$. As $\mu_{c_N}$ is supported 
by $[\sigma^2 (1-\sqrt{c_N})^2, \sigma^2 (1+\sqrt{c_N})^2]$, the almost sure convergence 
(\ref{eq:convergence-MP}) holds not only on $\mathbb{C} - \mathbb{R}^{+}$, but
also for each $z \in \mathbb{C} - \mathcal{I}_{\epsilon}$. In particular, 
(\ref{eq:convergence-MP}) is valid for $z=0$. Solving the equation (\ref{eq:stieltjes-MP}) for $z=0$
leads immediately to $m_{c_N}(0) = \frac{1}{\sigma^{2}(1-c_N)}$, and to 
\begin{equation}
\label{eq:comportement-as-trace-Q}
\lim_{N \rightarrow +\infty} \frac{1}{M} \mathrm{Tr}\left( {\bs \Sigma}_N {\bs \Sigma}_N^{*} \right)^{-1}  - \frac{1}{\sigma^{2}(1-c_N)} = 0
\end{equation}
almost surely. Taking the derivative of (\ref{eq:convergence-MP}) w.r.t. $z$ at
$z=0$, and using that $m_{c_N}^{'}(0) = \frac{1}{\sigma^{4}(1-c_N)^{3}}$, we
also obtain that
\begin{equation}
\label{eq:comportement-as-trace-Q^2}
\lim_{N \rightarrow +\infty} \frac{1}{M} \mathrm{Tr}\left( {\bs \Sigma}_N {\bs \Sigma}_N^{*} \right)^{-2}  - 
\frac{1}{\sigma^{4}(1-c_N)^{3}} = 0
\end{equation}
almost surely. Moreover, it is possible to specify the convergence speed in
(\ref{eq:comportement-as-trace-Q}) and (\ref{eq:comportement-as-trace-Q^2}). The following proposition is a direct consequence of Theorem 9.10 in \cite{bai-silverstein-book}.
\begin{proposition}
\label{prop:convergence-speed-traceQ}
It holds that 
\begin{align}
\label{eq:convergence-speed-trace-Q}
\frac{1}{M} \mathrm{Tr}\left(  {\bs \Sigma}_N {\bs \Sigma}_N^{*} \right)^{-1} - \frac{1}{\sigma^{2} (1-c_N)} & = \mathcal{O}_P(\frac{1}{N}) \\ 
\label{eq:convergence-speed-trace-Q^2}
\frac{1}{M} \mathrm{Tr}\left(  {\bs \Sigma}_N {\bs \Sigma}_N^{*}  \right)^{-2} - \frac{1}{\sigma^{4} (1-c_N)^{3}} & = \mathcal{O}_P(\frac{1}{N})
\end{align}
\end{proposition}
Theorem 9.10 in \cite{bai-silverstein-book} implies that the left hand side of 
(\ref{eq:convergence-speed-trace-Q}), renormalized by $N$, converges in distribution towards a Gaussian distribution, which, 
in turn, leads to (\ref{eq:convergence-speed-trace-Q}). (\ref{eq:convergence-speed-trace-Q^2}) holds for the same reason. 
\begin{remark}
\label{re:terme-de-moyenne}
As $c_N \rightarrow c$, the previous results of course imply that $\frac{1}{M} \mathrm{Tr}\left(  {\bs \Sigma}_N {\bs \Sigma}_N^{*} \right)^{-1}$ (resp. $\frac{1}{M} \mathrm{Tr}\left(  {\bs \Sigma}_N {\bs \Sigma}_N^{*}  \right)^{-2}$)  converge towards $\frac{1}{\sigma^{2} (1-c)}$ (resp. $\frac{1}{\sigma^{4} (1-c)^{3}}$). 
However, the rate of convergence is not a $\mathcal{O}_P(\frac{1}{N})$ term if 
the convergence speed of $c_N$ towards $c$ is less than $\mathcal{O}(\frac{1}{N})$. Therefore, 
it is more relevant to approximate the left hand sides of
(\ref{eq:convergence-speed-trace-Q}) and (\ref{eq:convergence-speed-trace-Q^2}) by $\frac{1}{\sigma^{2} (1-c_N)}$ and $\frac{1}{\sigma^{4} (1-c_N)^{3}}$. 
\end{remark}
The above results allow to characterize the asymptotic behaviour of 
the normalized trace of $\left( {\bs \Sigma}_N {\bs \Sigma}_N^{*} \right)^{-1}$
and $\left( {\bs \Sigma}_N {\bs \Sigma}_N^{*} \right)^{-2}$. However, it is also useful to 
obtain similar results on the bilinear forms of these matrices. 
\begin{proposition}
\label{prop:comportement-formes-quadratiques}
We consider two deterministic $M$-dimensional unit norm vectors ${\bf u}_N$ and
${\bf v}_N$. Then, it holds that 
\begin{equation}
\label{eq:comportement-as-forme-quadratique-Q}
\lim_{N \rightarrow +\infty} {\bf u}_N^* \left( {\bs \Sigma}_N {\bs \Sigma}_N^{*} \right)^{-1} {\bf v}_N   - \frac{{\bf u}_N^* {\bf v}_N}{\sigma^{2}(1-c_N)} = 0
\end{equation}
and that
\begin{equation}
\label{eq:comportement-as-forme-quadratique-Q^2}
\lim_{N \rightarrow +\infty} {\bf u}_N^* \left( {\bs \Sigma}_N {\bs \Sigma}_N^{*} \right)^{-2} {\bf v}_N   - \frac{{\bf u}_N^* {\bf v}_N}{\sigma^{4}(1-c_N)^{3}} = 0
\end{equation}
almost surely. Moreover, 
\begin{equation}
\label{eq:convergence-speed-forme-quadratique-Q}
 {\bf u}_N^* \left( {\bs \Sigma}_N {\bs \Sigma}_N^{*} \right)^{-1} {\bf v}_N   - \frac{{\bf u}_N^* {\bf v}_N}{\sigma^{2}(1-c_N)} = \mathcal{O}_P(\frac{1}{\sqrt{N}})
\end{equation}
Finally, if ${\bf C}_N$ is a positive $M \times M$ matrix such that 
$\mathrm{Rank}({\bf C}_N) = K$ is independent of $N$, and satisfying for each $N$
$0 < d_1 \leq \mathrm{Tr}({\bf C}_N^{2}) < d_2 < \infty$ for some constants
$d_1$ and $d_2$, then, we consider the sequence of random variables $(\kappa_N)_{N \geq 1}$
defined by
\begin{equation}
\label{eq:def-kappa}
\kappa_N = \mathrm{Tr}\left( {\bf C}_N ({\bs \Sigma}_N {\bs \Sigma}_N^{*})^{-1} \right)  
\end{equation}
Define by $\theta_N$ the term
\begin{equation}
\label{eq:def-theta}
\theta_N = \frac{\mathrm{Tr}({\bf C}_N^{2})}{\sigma^{4}(1-c_N)^{3}}
\end{equation}
Then, it holds that
\begin{multline}
\label{eq:convergence-fonction-caracteristique}
\mathbb{E}\left[ \exp\left( i u \sqrt{N} \left(\kappa_N - \frac{\mathrm{Tr}({\bf
C}_N)}{\sigma^{2}(1 - c_N)} \right) \right) \right] \\ - \exp \left(-
\frac{\theta_N u^{2}}{2}\right) \rightarrow 0
\end{multline}
for each $u \in \mathbb{R}$, and that 
\begin{equation}
\label{eq:tlc-forme-quadratique}
\frac{\sqrt{N}}{\sqrt{\theta_N}} \,  \left( \kappa_N - \frac{\mathrm{Tr}({\bf C}_N)}{\sigma^{2}(1 - c_N)} \right) \rightarrow_{\mathcal D} \; \mathcal{N}(0,1) 
\end{equation}
\end{proposition}
The almost sure convergence result (\ref{eq:comportement-as-forme-quadratique-Q}) is well known (see e.g. \cite{bilinear-ihp} in the context of a more general matrix model), while (\ref{eq:comportement-as-forme-quadratique-Q^2}) 
can be established by differentiating the behaviour of the bilinear forms of
${\bf Q}_N(z)$ w.r.t. $z$. Moreover,
(\ref{eq:convergence-speed-forme-quadratique-Q}) is a consequence of (\ref{eq:tlc-forme-quadratique}) used for the rank 1 matrix ${\bf C}_N = {\bf v}_N {\bf u}_N^{*}$.
(\ref{eq:convergence-fonction-caracteristique}) 
and (\ref{eq:tlc-forme-quadratique}) are new and need to be established.

A technical difficulty appears in the present context because we consider the
resolvent of the matrix ${\bs \Sigma}_N {\bs \Sigma}_N^{*}$ at $z=0$ while in
previous works, $z$ is supposed to belong to $\mathbb{C} - \mathbb{R}^{+}$.
To solve this issue, we use the regularization technic introduced in a more
general context in \cite{hac-lou-mes-naj-val-2011-rmta}.
For the proof, we refer the reader to Appendix
\ref{appendix:proof-of-quadratic form}.

We finish this appendix by a standard result whose proof is omitted. 
\begin{proposition}
\label{prop:trace-lemma}
We consider a $M \times L$ random matrix ${\bs \Gamma}_N$ with $\mathcal{N}_{\mathbb{C}}(0, \frac{\sigma^{2}}{N})$  i.i.d. entries, as well as the following deterministic matrices: 
${\bf A}_N$ is $M \times M$ and hermitian, ${\bf B}_N$ is $M \times L$ 
and satisfies $\sup_{N} \| {\bf B}_N \| < +\infty$ while ${\bf D}_N$ is a positive $L \times L$ matrix and also 
verifies  $\sup_{N} \| {\bf D}_N \| < +\infty$. Then, if $(\omega_N)_{N \geq 1}$
represents the sequence of random variables defined by 
\begin{equation}
\label{eq:def-omega}
\omega_N =  \mathrm{Tr} \left[ {\bf D}_N \left( {\bs \Gamma}_N^{*} {\bf A}_N {\bs \Gamma}_N +
{\bs \Gamma}_N^{*} {\bf B}_N + {\bf B}_N^{*} {\bs \Gamma}_N \right) \right]
\end{equation}
it holds that 
\begin{equation}
\label{eq:moyenne-omega}
\mathbb{E}(\omega_N)  =  \sigma^{2} \frac{1}{N} \mathrm{Tr}({\bf A}_N) \, \mathrm{Tr}({\bf D}_N) , 
\end{equation}
$$
\mathrm{Var}(\omega_N)  = \frac{1}{N} \zeta_N
$$
where $\zeta_N$ is defined by 
\begin{align}
\label{eq:variance-omega}
\zeta_N =& \sigma^{4} \,  \frac{1}{N} \mathrm{Tr}({\bf A}_N^{2} )
 \mathrm{Tr}({\bf D}_N^{2}) + 2 \sigma^{2} \, \frac{1}{N} \mathrm{Tr}\left(
 {\bf D}_N^{2} {\bf B}_N^{*} {\bf B}_N \right)
\end{align}
Moreover, 
\begin{align*}
\label{eq:moment-ordre-4-omega}
\mathbb{E} \left| \omega_N - \mathbb{E}(\omega_N) \right|^{4} \leq &
\frac{a_1}{N^{2}} + \frac{a_2}{N^{2}} \left(\frac{1}{N} \mathrm{Tr}({\bf
A}_N^{2} ) \right)^{2} \\ &+ \frac{a_3}{N^{3}} \frac{1}{N} \mathrm{Tr}({\bf
A}_N^{8}) \numberthis
\end{align*}
where $a_1, a_2, a_3$ are constant terms depending on $L, \sup_{N} \| {\bf B}_N \|$ and $\sup_{N} \| {\bf D}_N \|$. Finally, if $\limsup_N \zeta_N < +\infty$, it holds that 
\begin{equation}
\label{eq:fonction-caracteristique-omega}
\mathbb{E} \left( \exp{i \,u \sqrt{N}\left(\omega_N - \mathbb{E}(\omega_N)\right)} \right) - e^{-\frac{u^{2} \zeta_N}{2}} \rightarrow 0
\end{equation}
for each $u \in \mathbb{R}$. 
\end{proposition}

\section{Proofs of Theorems \ref{theo:eta0} and \ref{theo:eta1}}
\label{appendix:proof-of-theorems}
{\bf Proof of Theorem \ref{theo:eta0}.} In order to establish Theorem \ref{theo:eta0}, we use the results of 
Appendix \ref{appendix:useful-technical-results} for the matrix ${\bs \Sigma}_N
= \frac{1}{\sqrt{N}} {\bf V}_1$. We note that $\frac{1}{\sqrt{N}} {\bf V}_1$ is a $M \times (N-L)$ 
matrix while the results of Appendix \ref{appendix:useful-technical-results} 
have been presented in the context of a $M \times N$ 
matrix. In principle, it should be necessary to exchange $N$ by $N-L$ in 
Propositions \ref{prop:convergence-extreme-eigenvalues} to 
\ref{prop:comportement-formes-quadratiques}. However, $c_N - \frac{M}{N-L} = \mathcal{O}(\frac{1}{N})$, so that it possible to use the results of the above propositions without exchanging 
$N$ by $N-L$. 

We first verify (\ref{eq:convergence-as-eta0}). For this, we introduce the 
event $\mathcal{E}_N$ defined by (\ref{eq:def-E}). We first remark that 
$\eta_{N} - \eta_{N} \, \mathbb{1}_{\mathcal{E}_N^{c}} \rightarrow 0, a.s$. It
is thus sufficient to study the behaviour of $\eta_{N} \, \mathbb{1}_{\mathcal{E}_N^{c}}$ which is also equal to 
\begin{equation}
\label{eq:expre-eta-regularise}
\eta_{N} \, \mathbb{1}_{\mathcal{E}_N^{c}} = \log \mathrm{det} \left( {\bf I} + {\bf F}_N \mathbb{1}_{\mathcal{E}_N^{c}} \right)
\end{equation}
We now study the behaviour of each entry $(k,l)$ of matrix $\mathbb{1}_{\mathcal{E}_N^{c}} {\bf F}_N$. 
For this, we use Proposition \ref{prop:trace-lemma} for ${\bf D}_N = {\bf e}_k {\bf e}_l^{T}$, 
${\bs \Gamma}_N = \frac{{\bf V}_2}{\sqrt{N}}$ and ${\bf A}_N = \mathbb{1}_{\mathcal{E}_N^{c}} 
\left( \frac{{\bf V}_1 {\bf V}_1^{*}}{N} \right)^{-1}$. ${\bf A}_N$ is of course not 
deterministic, but as ${\bf V}_2$ and ${\bf V}_1$ are independent, it is possible to 
use the results of Proposition \ref{prop:trace-lemma} by replacing the mathematical 
expectation operator by the mathematical expectation operator $\mathbb{E}_{{\bf V}_2}$ 
w.r.t. ${\bf V}_2$. We note that 
the present matrix ${\bf A}_N$ verifies 
\begin{equation}
\label{eq:A-regularisee-borne}
{\bf A}_N \leq \frac{{\bf I}}{\sigma^{2}(1 - \sqrt{c})^{2} - \epsilon}
\end{equation}
because $\mathbb{1}_{\mathcal{E}_N^{c}} \neq 0$ implies that all the eigenvalues 
of $ \frac{{\bf V}_1 {\bf V}_1^{*}}{N} $ belong to 
$\mathcal{I}_{\epsilon} = [\sigma^{2}(1 - \sqrt{c})^{2} - \epsilon, \sigma^{2}(1 + \sqrt{c})^{2} + \epsilon]$. Therefore, (\ref{eq:moment-ordre-4-omega}) immediately implies that
\begin{align}
\mathbb{E}_{{\bf V}_2} \left| {\bf F}_{N,k,l} \mathbb{1}_{\mathcal{E}_N^{c}} - \mathbb{E}_{{\bf V}_2} \left({\bf F}_{N,k,l} \mathbb{1}_{\mathcal{E}_N^{c}} \right) \right|^{4} 
\leq \frac{a}{N^{2}}
\end{align}
where $a$ is a deterministic constant. Taking the mathematical 
expectation of the above inequality w.r.t. ${\bf V}_1$, 
and using the Borel-Cantelli Lemma lead to 
\begin{align}
{\bf F}_{N,k,l} \mathbb{1}_{\mathcal{E}_N^{c}} - \mathbb{E}_{{\bf V}_2} \left({\bf F}_{N,k,l} 
\mathbb{1}_{\mathcal{E}_N^{c}}  \right) \rightarrow 0 \; a.s.
\end{align}
or equivalently, to
\begin{align}
{\bf F}_{N,k,l} \mathbb{1}_{\mathcal{E}_N^{c}} - \delta(k-l) \, \sigma^{2} c_N  \, \frac{1}{M} \mathrm{Tr}\left( \frac{{\bf V}_1 {\bf V}_1^{*}}{N} \right)^{-1}  \rightarrow 0 \; a.s.
\end{align}
(\ref{eq:comportement-as-trace-Q}) implies that ${\bf F}_{N,k,l} \mathbb{1}_{\mathcal{E}_N^{c}} - \delta(k-l) \,\frac{c_N}{1-c_N} \rightarrow 0$ almost surely, or equivalently that 
\begin{align}
{\bf F}_N - \frac{c_N}{1-c_N} {\bf I} \rightarrow 0 \; a.s.
\end{align}
This eventually leads to (\ref{eq:convergence-as-eta0}). 

We now establish (\ref{eq:tlc-HO}). For this, we first remark that (\ref{eq:evaluation-indicatrice-Ec}) 
implies that $\eta_N = \eta_N \mathbb{1}_{\mathcal{E}_N^{c}} + \mathcal{O}_{P}(\frac{1}{N^{p}})$ 
for each integer $p$. Therefore, the asymptotic behaviour of the distribution 
of the left hand side of (\ref{eq:tlc-HO}) is not modified if $\eta_N$ is
replaced by $\eta_N \mathbb{1}_{\mathcal{E}_N^{c}}$ given by (\ref{eq:expre-eta-regularise}). 
We denote by ${\bs \Delta}_N$ the matrix defined by 
\begin{align}
{\bs \Delta}_N = {\bf F}_N \mathbb{1}_{\mathcal{E}_N^{c}} - \frac{c_N}{1-c_N} {\bf I}
\end{align}
We first prove that ${\bs \Delta}_N = \mathcal{O}_P(\frac{1}{\sqrt{N}})$. For this, 
we express ${\bs \Delta}_N$ as 
\begin{multline}
\label{eq:expre-Delta}
{\bs \Delta}_N = \left( {\bf F}_N \mathbb{1}_{\mathcal{E}_N^{c}} -  \sigma^{2} c_N  \, \frac{1}{M} \mathrm{Tr}\left( \frac{{\bf V}_1 {\bf V}_1^{*}}{N} \right)^{-1} \mathbb{1}_{\mathcal{E}_N^{c}}  {\bf I} \right)  \; + \\
\sigma^{2} c_N  \, \frac{1}{M} \mathrm{Tr}\left( \frac{{\bf V}_1 {\bf V}_1^{*}}{N} \right)^{-1} \mathbb{1}_{\mathcal{E}_N^{c}}  {\bf I}
- \frac{c_N}{1-c_N} {\bf I} 
\end{multline}
The first term of the right hand side of (\ref{eq:expre-Delta}) is
$\mathcal{O}_P(\frac{1}{\sqrt{N}})$ because the fourth-order moments of its entries are $\mathcal{O}(\frac{1}{N^{2}})$ terms. 
As for the second term, (\ref{eq:convergence-speed-trace-Q}) implies that it is a $\mathcal{O}_P(\frac{1}{N})$. A standard second order expansion of $\log \mathrm{det}( {\bf I} + {\bf F}_N \mathbb{1}_{\mathcal{E}_N^{c}})$ leads to 
\begin{align}
 \eta_N \mathbb{1}_{\mathcal{E}_N^{c}} = L \log \frac{1}{1 - c_N} + (1 - c_N) \mathrm{Tr}({\bs \Delta}_N) + \mathcal{O}_P(\frac{1}{N})
\end{align}
Therefore, it holds that 
\begin{align*}
\sqrt{N} \left( \eta_N \mathbb{1}_{\mathcal{E}_N^{c}} - L \log \frac{1}{1 - c_N} \right) 
=\sqrt{N}& (1 - c_N) \mathrm{Tr}({\bs \Delta}_N)\\
+& \mathcal{O}_P(\frac{1}{\sqrt{N}}), \numberthis
\end{align*}
or, using (\ref{eq:expre-Delta}), that
\begin{align*}
\sqrt{N} \left( \eta_N \mathbb{1}_{\mathcal{E}_N^{c}} - L \log \frac{1}{1 - c_N}\right) &= \\
\sqrt{N} (1 - c_N) \mathrm{Tr}\Big( 
{\bf F}_N  \mathbb{1}_{\mathcal{E}_N^{c}} - \sigma^{2} c_N & \, \frac{1}{M} 
\mathrm{Tr}\left( \frac{{\bf V}_1 {\bf V}_1^{*}}{N} \right)^{-1} 
\mathbb{1}_{\mathcal{E}_N^{c}} \Big) 
\\ + \mathcal{O}_P(\frac{1}{\sqrt{N}}) \numberthis
\end{align*}
As 
\begin{align}
\mathbb{E}_{{\bf V}_2}\left(  \mathrm{Tr}\left( 
{\bf F}_N \mathbb{1}_{\mathcal{E}_N^{c}} \right) \right) =  \sigma^{2} c_N  \, \frac{1}{M} \mathrm{Tr}\left( \left( \frac{{\bf V}_1 {\bf V}_1^{*}}{N} \right)^{-1} \mathbb{1}_{\mathcal{E}_N^{c}} \right),
\end{align}
Proposition \ref{prop:trace-lemma} used for ${\bf A}_N = \left( \frac{{\bf V}_1 {\bf V}_1^{*}}{N} \right)^{-1} \mathbb{1}_{\mathcal{E}_N^{c}}$, 
${\bf B}_N = 0$ and ${\bf D}_N = (1 - c_N) \mathbf{I}$
leads to 
\begin{multline}
\mathbb{E}_{{\bf V}_2} \left( \exp{ i u \sqrt{N}  \left( \eta_N - L \log\frac{1}{1 - c_N} \right)} \right) - \\
 \exp \left[-\frac{u^{2}}{2} \, \sigma^{4} \, L \, (1 - c_N)^{2} c_N  \, 
 \frac{1}{M} \mathrm{Tr}\left( \frac{{\bf V}_1 {\bf V}_1^{*}}{N} \right)^{-2} 
 \mathbb{1}_{\mathcal{E}_N^{c}} \right] \rightarrow 0
\end{multline}
a.s. for each $u \in \mathbb{R}$. (\ref{eq:comportement-as-trace-Q^2}) and the
dominated convergence theorem finally implies that 
\begin{multline}
\mathbb{E} \left( \exp{ i u \sqrt{N}  \left( \eta_N - L \log\frac{1}{1 - c_N} \right)} \right) - \\
 \exp \left[ - \frac{u^{2}}{2} \, \frac{L c_N}{1 - c_N} \right]  \rightarrow 0 
\end{multline}
This establishes (\ref{eq:tlc-HO}).

{\bf Proof of Theorem \ref{theo:eta1}}
We recall that, under $\mathrm{H_1}$, $\eta_N$ is given by (\ref{eq:expre-eta-H1}). 
As in the proof of Theorem  \ref{theo:eta0}, it is sufficient to study the regularized statistics 
$\eta_N \mathbb{1}_{\mathcal{E}_N^{c}}$ which is also equal to 
\begin{align}
\eta_N \mathbb{1}_{\mathcal{E}_N^{c}} = \log \mathrm{det} \left( {\bf I}_L +  \mathbb{1}_{\mathcal{E}_N^{c}} \, {\bf G}_N \right)
\end{align}
In order to evaluate the almost sure behaviour of $\eta_N \mathbb{1}_{\mathcal{E}_N^{c}}$, we expand  ${\bf G}_N  \mathbb{1}_{\mathcal{E}_N^{c}}$ as
\begin{multline}
\label{eq:expansion-G-appendix}
{\bf G}_N  \mathbb{1}_{\mathcal{E}_N^{c}} = {\bf H}^{*} \left( {\bf V}_1 {\bf V}_1^{*} / N \right)^{-1} {\bf H} \; \mathbb{1}_{\mathcal{E}_N^{c}}  + {\bf F}_N  \mathbb{1}_{\mathcal{E}_N^{c}} \, + \\
({\bf V}_2 / \sqrt{N})^{*} \left( {\bf V}_1 {\bf V}_1^{*} / N \right)^{-1} {\bf H} \; \mathbb{1}_{\mathcal{E}_N^{c}} \, + \\
{\bf H}^{*} \left( {\bf V}_1 {\bf V}_1^{*} / N \right)^{-1}  ({\bf V}_2 / \sqrt{N}) \; 
\mathbb{1}_{\mathcal{E}_N^{c}} 
\end{multline}
By (\ref{eq:comportement-as-forme-quadratique-Q}), the first term of the right
hand side of (\ref{eq:expansion-G-appendix}) behaves almost surely as $\frac{{\bf H}^{*} {\bf H}}{\sigma^{2}(1 - c_N)}$,
while it has been shown before that the second term converges a.s. 
towards $\frac{c_N}{1-c_N} {\bf I}$. To address the behaviour of entry $(k,l)$ of the sum of the third and the fourth terms, we use Proposition \ref{prop:trace-lemma} for ${\bs \Gamma}_N = \frac{{\bf V}_2}{\sqrt{N}}$, ${\bf A}_N = 0$, ${\bf B}_N = \left( {\bf V}_1 {\bf V}_1^{*} / N \right)^{-1} {\bf H} \; \mathbb{1}_{\mathcal{E}_N^{c}}$ and ${\bf D}_N = {\bf e}_k {\bf e}_l^{T}$. (\ref{eq:moment-ordre-4-omega})
implies that entry $(k,l)$ converges almost surely towards 0. Therefore, we have proved 
that 
\begin{align}
{\bf G}_N - \left( \frac{{\bf H}^{*} {\bf H}}{\sigma^{2}(1 - c_N)} +
\frac{c_N}{1-c_N} {\bf I} \right) \rightarrow \mathbf{0} \; a.s.
\end{align}
from which (\ref{eq:convergence-as-eta1}) follows immediately. 

The proof of (\ref{eq:tlc-H1}) is similar to the proof of (\ref{eq:tlc-HO}),
thus we do not provide all the details. 
We replace $\eta_N$ by $\eta_N \mathbb{1}_{\mathcal{E}_N^{c}}$, and remark that 
the matrix ${\bs \Delta}_N$, given by
\begin{align}
{\bs \Delta}_N = {\bf G}_N \mathbb{1}_{\mathcal{E}_N^{c}} -  \left( \frac{{\bf H}^{*} {\bf H}}{\sigma^{2}(1 - c_N)} + \frac{c_N}{1-c_N} {\bf I} \right) 
\end{align}
verifies ${\bs \Delta}_N = \mathcal{O}_P(\frac{1}{\sqrt{N}})$. To check this, 
it is sufficient to use the expansion (\ref{eq:expansion-G}), and to recognize that:
\begin{itemize}
\item by (\ref{eq:convergence-speed-forme-quadratique-Q}), 
\begin{align}
{\bf H}^{*} \left( {\bf V}_1 {\bf V}_1^{*} / N \right)^{-1} {\bf H} \; \mathbb{1}_{\mathcal{E}_N^{c}} - \frac{{\bf H}^{*} {\bf H}}{\sigma^{2}(1 - c_N)} = \mathcal{O}_P(\frac{1}{\sqrt{N}}), 
\end{align}
\item by Proposition \ref{prop:trace-lemma} and (\ref{eq:moment-ordre-4-omega}), 
\begin{multline}
({\bf V}_2 / \sqrt{N})^{*} \left( {\bf V}_1 {\bf V}_1^{*} / N \right)^{-1} {\bf H} \; \mathbb{1}_{\mathcal{E}_N^{c}} \, + \\
{\bf H}^{*} \left( {\bf V}_1 {\bf V}_1^{*} / N \right)^{-1}  ({\bf V}_2 / \sqrt{N}) \; 
\mathbb{1}_{\mathcal{E}_N^{c}} = \mathcal{O}_P(\frac{1}{\sqrt{N}})
\end{multline}
\item it has been shown before that 
\begin{align}
{\bf F}_N \mathbb{1}_{\mathcal{E}_N^{c}} - \frac{c_N}{1-c_N} {\bf I} = \mathcal{O}_P(\frac{1}{\sqrt{N}}).
\end{align}
\end{itemize}
Using a standard linearization of $\log \mathrm{det}({\bf I} + {\bf G}_N \mathbb{1}_{\mathcal{E}_N^{c}})$, this implies that 
\begin{equation}
\label{eq:linearisation-sous-H1}
\eta_N \mathbb{1}_{\mathcal{E}_N^{c}} - \overline{\eta}_{N,1} =
\mathrm{Tr}\left( {\bf D}_N \boldsymbol{\Delta}_N \right) + \mathcal{O}_P(1/N)
\end{equation}
where  ${\bf D}_N$ is the $L \times L$ matrix given by
\begin{align}
{\bf D}_{N} =  (1 -c_N)  ({\bf I}_L + {\bf H}^{*}{\bf H}/\sigma^{2})^{-1}
\end{align}
We define $\kappa_N$ and $\omega_N$  by
\begin{align}
\kappa_N = \mathrm{Tr}\left( {\bf C}_{N} \left( {\bf V}_1 {\bf V}_1^{*} / N \right)^{-1} \right)
\end{align}
and 
\begin{multline}
\omega_N = \mathrm{Tr} \left[ {\bf D}_{N}  {\bf F}_N  \mathbb{1}_{\mathcal{E}_N^{c}} \right] + \\
\mathrm{Tr} \left[ {\bf D}_{N} 
({\bf V}_2 / \sqrt{N})^{*} \left( {\bf V}_1 {\bf V}_1^{*} / N \right)^{-1} {\bf H} \; \mathbb{1}_{\mathcal{E}_N^{c}} \right] \, + \\
\mathrm{Tr} \left[ {\bf D}_{N} \,  {\bf H}^{*} \left( {\bf V}_1 {\bf V}_1^{*} / N \right)^{-1}  ({\bf V}_2 / \sqrt{N}) \; 
\mathbb{1}_{\mathcal{E}_N^{c}} \right]
\end{multline}
where ${\bf C}_{N}$ the $M \times M$ matrix given by 
\begin{align}
{\bf C}_{N} =  (1 -c_N)  {\bf H} ({\bf I}_L + {\bf H}^{*}{\bf H}/\sigma^{2})^{-1}  {\bf H}^{*}
\end{align}
Using (\ref{eq:linearisation-sous-H1}), we obtain that 
\begin{eqnarray}
 \eta_N \mathbb{1}_{\mathcal{E}_N^{c}} - \overline{\eta}_{N,1}  =  \kappa_N -  \frac{\mathrm{Tr}({\bf C}_N)}{\sigma^{2}(1 - c_N)} \; + \\ 
\nonumber
  \omega_N - \frac{c_N}{1 - c_N} \mathrm{Tr}({\bf D}_{N}) + \mathcal{O}_P(\frac{1}{N})
\end{eqnarray}
We also remark that (\ref{eq:convergence-speed-trace-Q}) used for ${\boldsymbol \Sigma}_N = 
\frac{1}{\sqrt{N}} {\bf V}_1$ implies that
\begin{align}
\omega_N - \mathbb{E}_{{\bf V}_2}(\omega_N) = \omega_N - \frac{c_N}{1 - c_N} \mathrm{Tr}({\bf D}_{N}) + 
\mathcal{O}_P(\frac{1}{N})
\end{align}
Therefore, it holds that 
\begin{align}
\sqrt{N} \left( \eta_N \mathbb{1}_{\mathcal{E}_N^{c}} - \overline{\eta}_{N,1} \right) = \sqrt{N} \left( \mathrm{Tr}\left( {\bf D}_N {\bs \Delta}_N \right) \right)
\end{align}
can be written as
\begin{multline}
\sqrt{N} \left( \eta_N \mathbb{1}_{\mathcal{E}_N^{c}} - \overline{\eta}_{N,1} \right)
= \sqrt{N} \left( \kappa_N - \frac{\mathrm{Tr}({\bf C}_N)}{\sigma^{2}(1 - c_N)} \right) + \\
\sqrt{N} \left(\omega_N - \mathbb{E}_{{\bf V}_2}(\omega_N) \right)+ \mathcal{O}_P(\frac{1}{\sqrt{N}})
\end{multline}
We denote by $\zeta_N$ the term
\begin{multline}
\zeta_N = \sigma^{4} \frac{1}{N} \mathrm{Tr} \left( ({\bf V}_1 {\bf V}_1^{*} / N )^{-2} \; \mathbb{1}_{\mathcal{E}_N^{c}} \right) \, \mathrm{Tr}({\bf D}_{N}^{2}) + \\
2 \sigma^{2} \frac{1}{N} \mathrm{Tr}\left({\bf D}_{N}^{2} {\bf H}^{*} ({\bf V}_1 {\bf V}_1^{*} / N)^{-1} {\bf H} \; \mathbb{1}_{\mathcal{E}_N^{c}}  \right)
\end{multline}
We use Proposition \ref{prop:trace-lemma} and (\ref{eq:fonction-caracteristique-omega})
for ${\bs \Gamma}_N = {\bf V}_2 / \sqrt{N}$, ${\bf A}_N =  ({\bf V}_1 {\bf V}_1^{*} / N )^{-1} \; \mathbb{1}_{\mathcal{E}_N^{c}} $ and ${\bf B}_N = ({\bf V}_1 {\bf V}_1^{*} / N )^{-1} \, {\bf H} \; \mathbb{1}_{\mathcal{E}_N^{c}} $, and obtain that 
\begin{multline}
\label{eq:fonction-caracteristique-conditionnelle}
\mathbb{E}_{{\bf V}_2} \left[ \exp \left( i u \, \sqrt{N} \left( \eta_N \mathbb{1}_{\mathcal{E}_N^{c}} - \overline{\eta}_{N,1} \right) \right) \right]- \\
\exp \left( i u \, \sqrt{N} \left( \kappa_N 
- \frac{\mathrm{Tr}({\bf C}_N)}{\sigma^{2}(1 - c_N)} \right) \right) \, 
\exp( - \frac{u^{2}}{2} \zeta_N) \rightarrow 0 \;
\end{multline}	
 a.s. $\zeta_N$ has almost surely the same behaviour as $\zeta$ given by
\begin{align}
\zeta = \frac{c_N}{(1-c_N)^{3}} \mathrm{Tr}({\bf D}_N^{2}) + 2 \frac{c_N}{(1-c_N)} \mathrm{Tr}({\bf D}_N^{2} {\bf H}^{*} {\bf H})
\end{align}
which implies that 
\begin{align}
\exp( - \frac{u^{2}}{2} \zeta_N) - \exp( - \frac{u^{2}}{2} \zeta) \rightarrow 0 \; a.s.
\end{align}
Therefore, taking the mathematical expectation of (\ref{eq:fonction-caracteristique-conditionnelle}) w.r.t ${\bf V}_1$ and using the dominated convergence theorem as well as
(\ref{eq:convergence-fonction-caracteristique}), lead, after some calculations, to 
\begin{multline} 
\label{eq:the-end}
\mathbb{E} \left[ \exp \left( i u \, \sqrt{N} ( \eta_N - \overline{\eta}_{N,1}) 
\right) \right] \\ - \exp\left[ - \frac{u^{2}}{2}\left( \frac{Lc_N}{1 - c_N} +
\kappa_1 \right) \right] \rightarrow 0
\end{multline}
for each $u$. As $\inf_N ( \frac{Lc_N}{1 - c_N} + \kappa_1) > 0$, 
(\ref{eq:tlc-H1}) follows from (\ref{eq:the-end}) 
(see Proposition 6 in \cite{r-hac-kor-lou-naj-pas-08}). \\

\section{Proof of (\ref{eq:tlc-forme-quadratique})}
\label{appendix:proof-of-quadratic form}
To establish (\ref{eq:tlc-forme-quadratique}),
we follow the approach of \cite{r-hac-kor-lou-naj-pas-08} which is based on the joint use
of the integration by parts formula and of the Poincar\'e-Nash inequality (see section III-B of
 \cite{r-hac-kor-lou-naj-pas-08}). However, the approach of \cite{r-hac-kor-lou-naj-pas-08}
allows to manage functionals of the resolvent ${\bf Q}_N(z)$ for $z \in \mathbb{C} - \mathbb{R}^{+}$.
For this, the inequality $\| {\bf Q}_N(z) \| \leq \frac{1}{\mathrm{dist}(z, \mathbb{R}^{+})}$  plays a fundamental role. For $z=0$,
$\|{\bf Q}_N(0)\|$ coincides with $\frac{1}{\lambda_{1,N}}$ which is not upper-bounded
by a deterministic positive constant for $N$ greater than a non random integer. 
This issue was solved before using the regularization term $\mathbb{1}_{\mathcal{E}_N^{c}}$.
However, the use of 
the integration by parts formula and the Poincar\'e-Nash inequality needs to consider smooth enough functions of
$ {\bs \Sigma}_N$. Motivated by \cite{hac-lou-mes-naj-val-2011-rmta}, we consider the regularization term $\chi_N$ defined by
\begin{equation}
\label{eq:def-chi}
\chi_N = \mathrm{det} \left[ \phi \left({\bs \Sigma}_N  {\bs \Sigma}_N^{*} \right) \right]
\end{equation}
where $\phi$ is a smooth function such that
$$
\phi(\lambda)  =  1 \; \mbox{if $\lambda \in \mathcal{I}_{\epsilon} =  [\sigma^{2}(1 - \sqrt{c})^{2} -  \epsilon,
\sigma^{2}(1 + \sqrt{c})^{2} +  \epsilon]$}
$$
$$
\phi(\lambda)  =  0 \; \mbox{if $\lambda \in  [\sigma^{2}(1 - \sqrt{c})^{2} - 2 \epsilon,
\sigma^{2}(1 + \sqrt{c})^{2} + 2 \epsilon]^{c}$}
$$
$$
\phi  \in  [0,1] \; \mbox{elsewhere}
$$
In the following, we need to use the following property: for each
$\epsilon > 0$, it holds that
\begin{equation}
\label{eq:escape-probability}
P\left(\mathcal{E}_N  \right) = \mathcal{O}(\frac{1}{N^{p}})
\end{equation}
where $\mathcal{E}_N$ is defined by (\ref{eq:def-E}). Property (\ref{eq:escape-probability}) is not mentioned in Theorem 5.11 of \cite{bai-silverstein-book} which addresses the non Gaussian case.
However, (\ref{eq:escape-probability}) follows directly from Gaussian concentration arguments.

It is clear that
\begin{equation}
\label{eq:borne-Q-regularise}
\left( {\bs \Sigma}_N {\bs \Sigma}_N^{*} \right)^{-1} \chi_N \leq \frac{{\bf
I}}{\sigma^{2}((1 - \sqrt{c})^{2} - 2 \epsilon)}
\end{equation}
Lemma 3-9 of \cite{hac-lou-mes-naj-val-2011-rmta} also
implies that, considered as a function of the entries of ${\bs \Sigma}_N$, $\chi_N$ is
continuously differentiable. Moreover, it follows from Proposition \ref{prop:convergence-extreme-eigenvalues}  that almost surely, for
$N$ large enough, $\chi_N = 1$ and $\kappa_N = \kappa_N \chi_N$. Therefore, it holds that
$\kappa_N \chi_N = \kappa_N + \mathcal{O}_P(\frac{1}{N^{p}})$, and that
\begin{multline}
\label{eq:effet-regularisation-kappa}
\sqrt{N} \left( \kappa_N -  \frac{\mathrm{Tr}({\bf C}_N)}{\sigma^{2}(1 - c_N)}
\right) \\ = \sqrt{N} \left( \kappa_N \chi_N - \frac{\mathrm{Tr}({\bf
C}_N)}{\sigma^{2}(1 - c_N)} \right) + \, \mathcal{O}_P(\frac{1}{N^{p}})
\end{multline}
for each $p \in \mathbb{N}$.
In order to establish (\ref{eq:convergence-fonction-caracteristique}), it is thus
sufficient to prove that
\begin{multline}
\label{eq:convergence-fonction-caracteristique-non-centree-kappa-regularise}
\mathbb{E}\left[ \exp\left( i u \sqrt{N} \left(\kappa_N \chi_N - \frac{\mathrm{Tr}({\bf C}_N)}{\sigma^{2}(1 - c_N)}  \right) \right) \right]
\\ - \exp \left(- \frac{\theta_N u^{2}}{2}\right)
\rightarrow 0
\end{multline}
for each $u$. To obtain (\ref{eq:tlc-forme-quadratique}),
we remark that, as  $\inf_N \theta_N > 0$, it follows from (\ref{eq:convergence-fonction-caracteristique-non-centree-kappa-regularise}) that
$$
\frac{\sqrt{N}}{\sqrt{\theta_N}}  \left(\kappa_N \chi_N - \frac{\mathrm{Tr}({\bf C}_N)}{\sigma^{2}(1 - c_N)}  \right) \rightarrow_{\mathcal{D}} \mathcal{N}_{\mathbb{R}}(0,1)
$$
(see Proposition 6 in \cite{r-hac-kor-lou-naj-pas-08}). (\ref{eq:tlc-forme-quadratique}) eventually appears as a consequence of (\ref{eq:effet-regularisation-kappa}).

The above regularization trick thus allows to replace the matrix $\left( {\bs
\Sigma}_N {\bs \Sigma}_N^{*} \right)^{-1}$ by $\left( {\bs \Sigma}_N {\bs \Sigma}_N^{*} \right)^{-1} \chi_N$, which verifies (\ref{eq:borne-Q-regularise}). In order to establish (\ref{eq:convergence-fonction-caracteristique-non-centree-kappa-regularise}), it is sufficient to prove that
\begin{equation}
\label{eq:convergence-moyenne-kappa-regularise}
\mathbb{E}(\kappa_N \chi_N) - \frac{\mathrm{Tr}({\bf C}_N)}{\sigma^{2}(1 - c_N)} = \mathrm{o}(\frac{1}{\sqrt{N}})
\end{equation}
and that
\begin{multline}
\label{eq:convergence-fonction-caracteristique-kappa-regularise}
\mathbb{E}\left[ \exp\left( i u \sqrt{N} \left(\kappa_N \chi_N -
\mathbb{E}(\kappa_N \chi_N)\right) \right) \right] \\
- \exp \left(-\frac{\theta_N u^{2}}{2}\right) \rightarrow 0
\end{multline}
for each $u$.

In the rest of this section, to
simplify the notations,
we omit to write the dependance on $N$ of the various terms
${\bs \Sigma}_N$, ${\bf Q}_N(0)$, $\chi_N$..., and denote them by
${\bs \Sigma}, {\bf Q}(0), \chi, \ldots$.
However, we keep the notation $c_N$, in order to avoid confusion between
$c_N$ and $c$. Furthermore, the matrix ${\bf Q}(0)$ is denoted
by ${\bf Q}$. If $x$ is a random variable, $x^{\circ}$ represents
the zero mean variable $x^{\circ} = x - \mathbb{E}(x)$. In the following,
we denote by $\delta$ the random variable defined by
$$
\delta = \sqrt{N} \, \kappa \, \chi
$$
and by $\psi^{\circ}(u)$ the characteristic function of $\delta^{\circ}$ defined by
$$
\psi^{\circ}(u) = \mathbb{E}\left( \exp{i u \delta^{\circ}} \right)
$$
We first establish the following Proposition.
\begin{proposition}
\label{prop:equa-diff-perturbee}
It holds that
\begin{equation}
\label{eq:equa-diff-approchee}
\left(\psi^{\circ}(u)\right)^{'} = -u \; \mathbb{E}\left(\mathrm{Tr}({\bf C}^{2} {\bf Q}^{2} \chi)\right) \, \psi^{\circ}(u)
+ \mathcal{O}(\frac{1}{\sqrt{N}})
\end{equation}
where $'$ represents the derivative w.r.t. the variable $u$.
\end{proposition}
{\bf Proof.} We consider the characteristic function $\psi(u)$ of $\delta$, and evaluate
$$
\psi^{'}(u) = i \sqrt{N} \mathbb{E}\left( \mathrm{Tr}({\bf Q} {\bf C} \chi) e^{i u \delta} \right)
$$
We remark that ${\bf Q} {\bs \Sigma} {\bs \Sigma}^{*} = {\bf I}$ so that
$$
\mathbb{E}\left( {\bf Q} {\bs \Sigma} {\bs \Sigma}^{*} \chi e^{i u \delta} \right)
= \mathbb{E}(  \chi e^{i u \delta})  {\bf I}
$$
We claim that
\begin{equation}
\label{eq:fonction-carac-regularisee}
\mathbb{E}(  \chi \, e^{i u \delta}) = \psi(u) + \mathcal{O}(\frac{1}{N^{p}})
\end{equation}
for each $p$. We remark that
$$
\left| \mathbb{E}\left(  e^{i u \delta} (1 - \chi) \right) \right| \leq 1 - \mathbb{E}(\chi)
$$
We recall that the event $\mathcal{E}$ is defined by (\ref{eq:def-E}) and that
$P(\mathcal{E}) =  \mathcal{O}(\frac{1}{N^{p}})$ for each $p$ (see
(\ref{eq:escape-probability})). $\mathbb{I}_{\mathcal{E}^{c}} \leq \chi$ leads to
$ 1 - \mathbb{E}(\chi)  \leq P(\mathcal{E})$. This justifies (\ref{eq:fonction-carac-regularisee}).
Therefore, it holds that
\begin{equation}
\label{eq:equation-resolvente-regularisee}
\mathbb{E}\left( {\bf Q} {\bs \Sigma} {\bs \Sigma}^{*} \chi \, e^{i u \delta} \right) =
\left(\psi(u) + \mathcal{O}(\frac{1}{N^{p}})\right) {\bf I}
\end{equation}
for each $p$. We now evaluate each entry of the lefthandside of (\ref{eq:equation-resolvente-regularisee})
using the integration by parts formula. For this, we denote by $({\bs \xi}_1, \ldots, {\bs \xi}_N)$ the columns
of ${\bs \Sigma}$. It holds that
$$
\left({\bf Q} {\bs \Sigma} {\bs \Sigma}^{*} \right)_{r,s}= \sum_{j=1}^{N} ({\bf Q} {\bs \xi_j})_{r}
\overline{{\bs \Sigma}}_{s,j}
$$
and that
$$
\mathbb{E}\left[ ({\bf Q} {\bs \xi_j})_{r}
\overline{{\bs \Sigma}}_{s,j} \chi \, e^{i u \delta} \right] =
\sum_{t=1}^{M} \mathbb{E}\left( {\bf Q}_{r,t} {\bs \Sigma}_{t,j} \overline{{\bs \Sigma}}_{s,j} \chi \, e^{i u \delta} \right)
$$
The integration by parts formula leads to
$$
\mathbb{E}\left( {\bf Q}_{r,t}  \overline{{\bs \Sigma}}_{s,j} \chi \, e^{i u \delta} {\bs \Sigma}_{t,j} \right) =
\frac{\sigma^{2}}{N} \mathbb{E} \left[ \frac{\partial \left({\bf Q}_{r,t}  \overline{{\bs \Sigma}}_{s,j} \chi \, e^{i u \delta}\right)}{\partial \overline{{\bs \Sigma}}_{t,j}} \right]
$$
After some algebra, we obtain that
\begin{multline}
\label{eq:first-step-novikov}
\mathbb{E}\left( {\bf Q}_{r,t}  \overline{{\bs \Sigma}}_{s,j} \chi \, e^{i u \delta} {\bs \Sigma}_{t,j} \right) =
\frac{\sigma^{2}}{N} \mathbb{E}\left( {\bf Q}_{r,t} \chi \, e^{i u \delta}  \right) \, \delta(t=s)  \\
- \, \frac{\sigma^{2}}{N} \mathbb{E}\left( ({\bf Q} {\bs \xi}_j)_{r} \, {\bf Q}_{t,t} \, \overline{{\bs \Sigma}}_{s,j} \, \chi \, e^{i u \delta} \right) \\
- \, \frac{i \sigma^{2} u}{\sqrt{N}} \mathbb{E} \left(  {\bf Q}_{r,t} \, ({\bf Q} {\bf C} {\bf Q}  {\bs \xi}_j)_{t}  \, \overline{{\bs \Sigma}}_{s,j} \, \chi \,  e^{i u \delta} \right)  \\
+ \,\frac{\sigma^{2}}{N} \mathbb{E}\left( {\bf Q}_{r,t}  \, \overline{{\bs \Sigma}}_{s,j}  \, e^{i u \delta}\, \frac{\partial \chi}{\partial \overline{{\bs \Sigma}}_{t,j}} \right)
\end{multline}
We now need to study more precisely the properties of the derivative of $\chi$
w.r.t. $ \overline{{\bs \Sigma}}_{t,j}$. For this, we give the following Lemma
\begin{lemma}
\label{le:proprietes-derivees-chi}
We denote by $\mathcal{A}$ the event:
\begin{multline}
\label{eq:def-A}
\mathcal{A} = \{ \mbox{one of the $\hat{\lambda}_{k,N}$ escapes from
$\mathcal{I}_{\epsilon}$} \} \\
\cap \{ (\hat{\lambda}_{l,N})_{l=1, \ldots,M}
\in
\mathrm{supp}(\phi) \}
\end{multline}
Then, it holds that
\begin{equation}
\label{eq:nullite-derivee-chi}
\frac{\partial \chi}{\partial \overline{{\bs \Sigma}}_{t,j}}=0 \; \mbox{on $\mathcal{A}^{c}$}
\end{equation}
and that
\begin{equation}
\label{eq:borne-derivee-chi}
\mathbb{E}   \left| \frac{\partial \chi}{\partial {\overline{\bs \Sigma}}_{i,j}} \right|^{2}  =   \mathcal{O}(\frac{1}{N^{p}})
\end{equation}
for each $p$.
\end{lemma}
{\bf Proof.}
Lemma \ref{le:proprietes-derivees-chi} follows directly from Lemma 3.9 of \cite{hac-lou-mes-naj-val-2011-rmta}
and from the calculations in the proof of Proposition 3.3 of  \cite{hac-lou-mes-naj-val-2011-rmta}.

Lemma \ref{le:proprietes-derivees-chi} implies that the last term of (\ref{eq:first-step-novikov})
is $\mathcal{O}(\frac{1}{N^{p}})$ for each $p$. To check this, we remark that
$$
\mathbb{E}\left( {\bf Q}_{r,t}  \, \overline{{\bs \Sigma}}_{s,j}  \, e^{i u \delta}\, \frac{\partial \chi}{\partial \overline{{\bs \Sigma}}_{t,j}} \right) =
\mathbb{E}\left( {\bf Q}_{r,t}  \, \overline{{\bs \Sigma}}_{s,j}  \, e^{i u \delta}\, \mathbb{1}_{\mathcal{A}}\frac{\partial \chi}{\partial \overline{{\bs \Sigma}}_{t,j}} \right)
$$
The Schwartz inequality leads to
\begin{align*}
| \mathbb{E}\Big( {\bf Q}_{r,t}  \, \overline{{\bs \Sigma}}_{s,j}  \, e^{i
u \delta}\, & \mathbb{1}_{\mathcal{A}} \frac{\partial \chi}{\partial
\overline{{\bs \Sigma}}_{t,j}} \Big) |^{2} \\
&\leq \mathbb{E}\Big( |{\bf Q}_{r,t}  \, \overline{{\bs \Sigma}}_{s,j}|^{2}
\,
\mathbb{1}_{\mathcal{A}} \Big) \mathbb{E}   \left| \frac{\partial
\chi}{\partial {\overline{\bs \Sigma}}_{i,j}} \right|^{2}
\end{align*}
On event $\mathcal{A}$, all the eigenvalues of ${\bs \Sigma} {\bs \Sigma}^{*}$ belong to $[\sigma^{2}(1 - \sqrt{c})^{2} - 2 \epsilon,
\sigma^{2}(1 + \sqrt{c})^{2} + 2 \epsilon]$. Therefore, $|{\bf Q}_{r,t}  \, \mathbb{1}_{\mathcal{A}}$ is bounded and (\ref{eq:borne-derivee-chi}) implies that the last term of (\ref{eq:first-step-novikov}) is $\mathcal{O}(\frac{1}{N^{p}})$ for each $p$. Summing (\ref{eq:first-step-novikov})
over $t$, we obtain that
\begin{multline}
\label{eq:step-2-novikov}
\mathbb{E}\left( ({\bf Q} {\bs \xi}_j)_{r}  \overline{{\bs \Sigma}}_{s,j} \, \chi \, e^{i u \delta} \right) =
\frac{\sigma^{2}}{N} \mathbb{E}\left( {\bf Q}_{r,s} \chi \, e^{i u \delta}  \right)  \\
- \, \sigma^{2} c_N \, \mathbb{E}\left( \hat{m}(0) \, ({\bf Q} {\bs \xi}_j)_{r} \, \overline{{\bs \Sigma}}_{s,j} \, \chi \, e^{i u \delta} \right) \\
- \, \frac{i \sigma^{2} u}{\sqrt{N}} \mathbb{E} \left( ({\bf Q}^{2} {\bf C} {\bf Q}  {\bs \xi}_j)_{r}  \, \overline{{\bs \Sigma}}_{s,j} \, \chi \,  e^{i u \delta} \right) +  \mathcal{O}(\frac{1}{N^{p}})
\end{multline}
where we recall that $\hat{m}(0) = \frac{1}{M} \mathrm{Tr}({\bf Q})$ represents the Stieltjes
transform of the empirical eigenvalue distribution $\hat{\mu}$ of ${\bs \Sigma} {\bs \Sigma}^{*}$ at
$z = 0$. Using that $(1 - \chi) \leq \mathbb{1}_{\mathcal{E}}$, it is easy to check that for each $p$,
it holds that
\begin{align*}
\mathbb{E}\Big( ({\bf Q} {\bs \xi}_j)_{r} \, \hat{m}(0) & \, \overline{{\bs
\Sigma}}_{s,j} \, \chi \, e^{i u \delta} \Big) = \\
&\mathbb{E}\left( ({\bf Q} {\bs \xi}_j)_{r} \,
\hat{m}(0) \, \overline{{\bs \Sigma}}_{s,j} \, \chi^{2} \, e^{i u \delta} \right)
 + \mathcal{O}(\frac{1}{N^{p}})
\end{align*}
We denote by $\beta$ the term $\beta = \hat{m}(0) \chi$, and express
$\beta$ as $\beta = \alpha + \beta^{\circ}$. Replacing $\chi$ by $\chi^{2}$ in the
second term of the righthandside of (\ref{eq:step-2-novikov}) and plugging $\beta = \alpha + \beta^{\circ}$
into (\ref{eq:step-2-novikov}), we obtain that immediately that
\begin{multline}
\label{eq:step-3-novikov}
\mathbb{E}\left( ({\bf Q} {\bs \xi}_j)_{r}  \overline{{\bs \Sigma}}_{s,j} \, \chi \, e^{i u \delta} \right) =
\frac{\sigma^{2}}{N(1 + \sigma^{2} c_N \alpha)} \mathbb{E}\left( {\bf Q}_{r,s} \chi \, e^{i u \delta}  \right)  \\
- \, \frac{i \sigma^{2} u}{\sqrt{N}(1 + \sigma^{2} c_N \alpha)} \mathbb{E} \left( ({\bf Q}^{2} {\bf C} {\bf Q}  {\bs \xi}_j)_{r}  \, \overline{{\bs \Sigma}}_{s,j} \, \chi \,  e^{i u \delta} \right) \\
- \, \frac{\sigma^{2} c_N}{1 + \sigma^{2} c_N \alpha} \mathbb{E}\left(  \beta^{\circ} \,  ({\bf Q} {\bs \xi}_j)_{r} \, \overline{{\bs \Sigma}}_{s,j} \, \chi \, e^{i u \delta} \right) +  \mathcal{O}(\frac{1}{N^{p}})
\end{multline}
Summing over $j$, we get that
\begin{multline}
\label{eq:step-4-novikov}
\mathbb{E}\left( ({\bf Q} {\bs \Sigma} {\bs \Sigma}^{*})_{r,s} \, \chi \, e^{i u \delta} \right) =
\frac{\sigma^{2}}{1 + \sigma^{2} c_N \alpha} \mathbb{E}\left( {\bf Q}_{r,s} \chi \, e^{i u \delta}  \right)  \\
- \, \frac{i \sigma^{2} u}{\sqrt{N}(1 + \sigma^{2} c_N \alpha)} \mathbb{E} \left( ({\bf Q}^{2} {\bf C} {\bf Q} {\bs \Sigma} {\bs \Sigma}^{*})_{r,s}  \, \chi \,  e^{i u \delta} \right) \\
- \, \frac{\sigma^{2} c_N}{1 + \sigma^{2} c_N \alpha} \mathbb{E}\left( \beta^{\circ}  ({\bf Q}  {\bs \Sigma} {\bs \Sigma}^{*})_{r,s} \, \chi \, e^{i u \delta} \right) +  \mathcal{O}(\frac{1}{N^{p}})
\end{multline}
or, using that ${\bf Q} {\bs \Sigma} {\bs \Sigma}^{*} = {\bf I}$,
\begin{multline}
\label{eq:step-5-novikov}
\mathbb{E}\left( \chi \, e^{i u \delta} \right) \, \delta(r=s) =
\frac{\sigma^{2}}{1 + \sigma^{2} c_N \alpha} \mathbb{E}\left( {\bf Q}_{r,s} \chi \, e^{i u \delta}  \right)  \\
- \, \frac{i \sigma^{2} u}{\sqrt{N}(1 + \sigma^{2} c_N \alpha)} \mathbb{E} \left( ({\bf Q}^{2} {\bf C})_{r,s}  \, \chi \,  e^{i u \delta} \right) \\
- \, \frac{\sigma^{2} c_N}{1 + \sigma^{2} c_N \alpha} \mathbb{E}\left( \beta^{\circ} \, \chi \, e^{i u \delta} \right) \, \delta(r=s) +  \mathcal{O}(\frac{1}{N^{p}})
\end{multline}
In order to evaluate $\alpha$, we take $u=0$ and sum over $r=s$ in (\ref{eq:step-5-novikov}),
and obtain that
$$
\alpha = \frac{1}{\sigma^{2}(1 - c_N)} + \frac{1}{1 - c_N} \mathbb{E}\left( \beta^{\circ} \, \chi \right)
+  \mathcal{O}(\frac{1}{N^{p}})
$$
$\mathbb{E}\left( \beta^{\circ} \, \chi \right)$ coincides with  $\mathbb{E}\left( \beta^{\circ} \, \chi^{\circ} \right)$. Using (\ref{eq:borne-derivee-chi}), the Poincar\'e-Nash inequality leads immediately
to $\mathbb{E}\left((\chi^{\circ})^{2}\right) = \mathcal{O}(\frac{1}{N^{p}})$, and to
\begin{equation}
\label{eq:expre-alpha}
\alpha = \frac{1}{\sigma^{2}(1 - c_N)} +  \mathcal{O}(\frac{1}{N^{p}})
\end{equation}
for each $p$. As a consequence, we also get that
\begin{equation}
\label{eq:expre-E(Q)-regularise}
\mathbb{E}({\bf Q}_{r,s} \, \chi) = \frac{1}{\sigma^{2}(1 - c_N)}  \, \delta(r=s) +  \mathcal{O}(\frac{1}{N^{p}})
\end{equation}
We now use (\ref{eq:step-5-novikov}) in order to evaluate $\mathbb{E}\left( ({\bf Q}_{r,s} \chi)^{\circ} \, \chi \, e^{i u \delta} \right)$. For this, we first establish  that the use of (\ref{eq:borne-Q-regularise}) and of the Poincar\'e-Nash inequality implies that
\begin{equation}
\label{eq:variance-beta}
\mathrm{Var}(\beta) = \mathbb{E}\left((\beta^{\circ})^{2}\right) = \mathcal{O}(\frac{1}{N^{2}})
\end{equation}
To check this, we use the Poincar\'e-Nash inequality:
$$
\mathrm{Var}(\beta) \leq \frac{\sigma^{2}}{N}   \mathbb{E} \left( \sum_{i,j}  \left| \frac{\partial \beta}{\partial {\bs \Sigma}_{i,j}} \right|^{2} + \left| \frac{\partial \beta}{\partial \overline{{\bs \Sigma}}_{i,j}} \right|^{2} \right)
$$
We just evaluate the terms corresponding to the derivatives with respect to the
terms $(\overline{{\bs \Sigma}}_{i,j})_{i=1, \ldots, M, j=1, \ldots, N}$. It is easily seen that
$$
\frac{\partial \beta}{\partial \overline{{\bs \Sigma}}_{i,j}} = -\frac{1}{M} ({\bf e}_i^{T} {\bf Q}^{2} {\bs \xi}_j) \, \chi + \frac{1}{M} \mathrm{Tr}({\bf Q}) \frac{\partial \chi}{\partial \overline{{\bs \Sigma}}_{i,j}}
$$
Therefore, it holds that
$$
\left| \frac{\partial \beta}{\partial \overline{{\bs \Sigma}}_{i,j}} \right|^{2} \leq
2 \frac{1}{M^{2}} {\bs \xi}_j^{*} {\bf Q}^{2} {\bf e}_i {\bf e}_i^{T} {\bf Q}^{2} {\bs \xi}_j \, \chi^{2}
+ 2  \frac{1}{M} \mathrm{Tr}({\bf Q}) \, \left| \frac{\partial \chi}{\partial \overline{{\bs \Sigma}}_{i,j}} \right|^{2}
$$
Using the identity ${\bf Q} {\bs \Sigma} {\bs \Sigma}^{*} = {\bf I}$ as well that
$\frac{\partial \chi}{\partial \overline{{\bs \Sigma}}_{i,j}} = \mathbb{1}_{\mathcal A} \frac{\partial \chi}{\partial \overline{{\bs \Sigma}}_{i,j}}$ (see (\ref{eq:nullite-derivee-chi})),
we obtain that
\begin{multline*}
\frac{\sigma^{2}}{N}  \sum_{i,j} \left( \mathbb{E} \left| \frac{\partial \beta}{\partial \overline{{\bs \Sigma}}_{i,j}} \right|^{2}  \right) \leq 2 \sigma^{2} \frac{1}{MN} \mathbb{E}\left(
\frac{1}{M} \mathrm{Tr}({\bf Q}^{3}) \, \chi \right) + \\
2 \frac{\sigma^{2}}{N} \mathbb{E} \left(  \frac{1}{M} \mathrm{Tr}({\bf Q}) \, \mathbb{1}_{\mathcal A}
\sum_{i,j} \left| \frac{\partial \chi}{\partial \overline{{\bs \Sigma}}_{i,j}} \right|^{2} \right)
\end{multline*}
On the set $\mathcal{A}$, the eigenvalues of $ {\bs \Sigma} {\bs \Sigma}^{*}$
are located into $[\sigma^{2}(1 - \sqrt{c})^{2} - 2 \epsilon, \sigma^{2}(1 + \sqrt{c})^{2} + 2 \epsilon]$.
Therefore, we get that
$$
\frac{1}{M} \mathrm{Tr}({\bf Q}) \, \mathbb{1}_{\mathcal A} \leq \frac{1}{\sigma^{2}(1 - \sqrt{c})^{2} - 2 \epsilon}
$$
Using (\ref{eq:borne-derivee-chi}), we obtain that
$$
2 \frac{\sigma^{2}}{N} \mathbb{E} \left(  \frac{1}{M} \mathrm{Tr}({\bf Q}) \, \mathbb{1}_{\mathcal A}
\sum_{i,j} \left| \frac{\partial \chi}{\partial \overline{{\bs \Sigma}}_{i,j}} \right|^{2} \right) =
\mathcal{O}(\frac{1}{N^{p}})
$$
for each $p$. Moreover, (\ref{eq:borne-Q-regularise}) implies that
$$
\frac{1}{M} \mathrm{Tr}({\bf Q}^{3}) \, \chi  \leq \frac{1}{\left(\sigma^{2}(1 - \sqrt{c})^{2} - 2 \epsilon\right)^{3}}
$$
and that
$$
2 \sigma^{2} \frac{1}{MN} \mathbb{E}\left(
\frac{1}{M} \mathrm{Tr}({\bf Q}^{3}) \, \chi \right)  = \mathcal{O}(\frac{1}{N^{2}})
$$
This establishes (\ref{eq:variance-beta}).

Therefore, the Schwartz inequality leads to $ \mathbb{E}\left( \beta^{\circ} \, \chi \, e^{i u \delta} \right) = \mathcal{O}(\frac{1}{N})$. Writing $ \mathbb{E}\left( {\bf Q}_{r,s} \chi \, e^{i u \delta}  \right) $ as
\begin{multline*}
\mathbb{E}\left( {\bf Q}_{r,s} \chi \, e^{i u \delta}  \right) = \mathbb{E}\left( {\bf Q}_{r,s} \chi^{2} \, e^{i u \delta}  \right) +  \mathcal{O}(\frac{1}{N^{p}}) = \\
\mathbb{E}({\bf Q}_{r,s} \chi) \mathbb{E}( \chi \, e^{i u \delta}) +
\mathbb{E}\left(({\bf Q}_{r,s} \chi)^{\circ} \chi \, e^{i u \delta} \right) + \mathcal{O}(\frac{1}{N^{p}}) = \\
\mathbb{E}({\bf Q}_{r,s} \chi) \mathbb{E}( \chi \, e^{i u \delta}) +
\mathbb{E}\left(({\bf Q}_{r,s} \chi)^{\circ} \, e^{i u \delta} \right) + \mathcal{O}(\frac{1}{N^{p}})
\end{multline*}
(\ref{eq:expre-alpha}), (\ref{eq:expre-E(Q)-regularise}) and (\ref{eq:step-5-novikov}) lead to
\begin{equation}
\label{eq:step-6-novikov}
\mathbb{E}\left(({\bf Q}_{r,s} \chi)^{\circ}  \, e^{i u \delta} \right) = \frac{i u}{\sqrt{N}} \,
\mathbb{E}\left( ({\bf Q}^{2} {\bf C})_{r,s} \, \chi \, e^{i u \delta} \right) +  \mathcal{O}(\frac{1}{N})
\end{equation}
or equivalently to
$$
\mathbb{E}\left( \delta^{\circ} \, e^{i u \delta} \right) = i u \, \mathbb{E}\left( \mathrm{Tr}({\bf Q}^{2} {\bf C}^{2}) \, \chi \, e^{i u \delta} \right) + \mathcal{O}(\frac{1}{\sqrt{N}})
$$
Using the Nash-Poincar\'e inequality, it can be checked that
$$
\mathrm{Var}\left( \mathrm{Tr}({\bf Q}^{2} {\bf C}^{2}) \, \chi \right) = \mathcal{O}(\frac{1}{N})
$$
Therefore, the Schwartz inequality leads to
$$
\mathbb{E}\left( \mathrm{Tr}({\bf Q}^{2} {\bf C}^{2}) \, \chi \, e^{i u \delta} \right) =
\mathbb{E}\left( \mathrm{Tr}({\bf Q}^{2} {\bf C}^{2}) \, \chi \right) \mathbb{E}( e^{i u \delta})
+  \mathcal{O}(\frac{1}{\sqrt{N}})
$$
and we get that
\begin{equation}
\label{eq:step-7-novikov}
\mathbb{E}\left( \delta^{\circ} \, e^{i u \delta} \right) = i u \, \mathbb{E}\left( \mathrm{Tr}({\bf Q}^{2} {\bf C}^{2}) \, \chi \right) \mathbb{E}( e^{i u \delta}) + \mathcal{O}(\frac{1}{\sqrt{N}})
\end{equation}
Plugging $\delta = \delta^{\circ} + \mathbb{E}(\delta)$ into (\ref{eq:step-7-novikov}) eventually leads to
\begin{equation}
\label{eq:step-8-novikov}
\mathbb{E}\left( \delta^{\circ} \, e^{i u \delta^{\circ}} \right) = i u \, \mathbb{E}\left( \mathrm{Tr}({\bf Q}^{2} {\bf C}^{2}) \, \chi \right) \mathbb{E}( e^{i u \delta^{\circ}}) + \mathcal{O}(\frac{1}{\sqrt{N}})
\end{equation}
which is equivalent to (\ref{eq:equa-diff-approchee}). This, in turn, establishes Proposition
\ref{prop:equa-diff-perturbee}. \\

We now complete the proof of (\ref{eq:convergence-fonction-caracteristique-kappa-regularise}). We integrate (\ref{eq:equa-diff-approchee}), and obtain that
$$
\psi^{\circ}(u) = \exp \left[ -\frac{u^{2}}{2} \, \mathbb{E}\left( \mathrm{Tr}({\bf Q}^{2} {\bf C}^{2} \, \chi) \right) \right] + \mathcal{O}(\frac{1}{\sqrt{N}})
$$
(see section V-C of \cite{r-hac-kor-lou-naj-pas-08} for more details). (\ref{eq:comportement-as-forme-quadratique-Q^2}) implies that
$$
\mathrm{Tr}({\bf Q}^{2} {\bf C}^{2}) - \frac{\mathrm{Tr}({\bf C}^{2})}{\sigma^{4}(1 - c_N)^{3}}
\rightarrow 0 \, a.s.
$$
As $\mathrm{Tr}({\bf Q}^{2} {\bf C}^{2}) \, \chi - \mathrm{Tr}({\bf Q}^{2} {\bf C}^{2})$ also
converges to $0$ almost surely, we obtain that
$$
\mathrm{Tr}({\bf Q}^{2} {\bf C}^{2}) \chi - \frac{\mathrm{Tr}({\bf C}^{2})}{\sigma^{4}(1 - c_N)^{3}}
\rightarrow 0 \, a.s.
$$
As matrix ${\bf Q}^{2} \chi$ is bounded and $\sup_N \mathrm{Tr}({\bf C}^{2}) < +\infty$,
it is possible to use the Lebesgue dominated convergence theorem and to conclude that
$$
\mathbb{E} \left(\mathrm{Tr}({\bf Q}^{2} {\bf C}^{2}) \chi \right) - \frac{\mathrm{Tr}({\bf C}^{2})}{\sigma^{4}(1 - c_N)^{3}} \rightarrow 0
$$
This proves (\ref{eq:convergence-fonction-caracteristique-kappa-regularise}).

It remains to establish (\ref{eq:convergence-moyenne-kappa-regularise}). For this,
we use (\ref{eq:expre-E(Q)-regularise}), and obtain that
$$
\mathbb{E}\left( \mathrm{Tr}({\bf Q} {\bf C}) \, \chi \right) - \frac{\mathrm{Tr}({\bf C})}{\sigma^{2}(1-c_N)} =
\mathcal{O}(\frac{1}{N^{p}})
$$
for each $p$. This, of course, implies
(\ref{eq:convergence-moyenne-kappa-regularise}).

\end{appendices}

\bibliographystyle{IEEEbib}

\begin{thebibliography}{99}

\bibitem{astely-jakobsson-1999} D. Astely, A. Jakobsson, A.L. Swindlehurst, 
"Burst synchronization on unknown frequency selective channels with 
co-channel interference using an antenna array", 
{\sl Proc. IEEE 49th Veh. Technol. Conf.} vol.49, pp.2363-2367, 1999

\bibitem{bai-silverstein-book} Z. Bai, J.W. Silverstein, 
"Spectral analysis of large dimensional random matrices", 
Springer Series in Statistics, 2nd ed., 2010. 

\bibitem{bianchi-et-al-2011} P. Bianchi, M. Debbah, M. Ma\"{e}da and J. Najim, 
"Performance of statistical tests for single source detection using random
matrix theory", {\sl IEEE Inf. Theory}, vol. 57, no.4, pp. 2400--2419, Apr. 2011  

\bibitem{bliss-2010} D. W. Bliss, P. A. Parker 
"Temporal synchronization of MIMO wireless communication in the presence of interference",
{\sl IEEE Trans. Signal Process.}, vol. 58, no.3, pp.1794-1806, Mar. 2010  

\bibitem{chu-1972} D. C. Chu, 
“Polyphase codes with good periodic correlation properties,”
{\sl IEEE Trans. Inf. Theory,}, vol. 18, no.4, pp. 531-532, Jul. 1972	  	

\bibitem{couillet-1} R. Couillet, J.W. Silverstein, Z.D. Bai, M.~Debbah, 
"Eigen-inference for energy estimation of multiple sources"
{\sl IEEE Trans. Inf. Theory}, vol. 57, no. 4, pp. 2420-2439, Apr. 2011.

\bibitem{dogandzic-nehorai-2002} A. Dogandzic, A. Nehorai, 
"Finite-length MIMO equalization using canonical correlation analysis", 
{\sl IEEE Trans. Signal Process.}, vol. 50, no. 4, pp. 984-989, Apr. 2002.

\bibitem{dumont-et-al-2010} J. Dumont, W. Hachem, S. Lasaulce, P. Loubaton, J. Najim, 
"On the capacity achieving transmit covariance matrices for MIMO Rician channels: 
an asymptotic approach", {\sl IEEE Transactions on Information Theory}, 
vol. 56, no. 3, pp. 1048-1069, Mar. 2010. 

\bibitem{girko-sara} V.L. Girko, "An introduction to statistical analysis of random arrays", 
VSP, The Netherlands, 1998.

\bibitem{r-hac-kor-lou-naj-pas-08} 
W. Hachem, O. Khorunzhiy, P. Loubaton, J. Najim, L. Pastur, 
"A new approach for mutual information analysis of large dimensional
multi-antenna channels", {\sl IEEE Trans. Inf. Theory}, vol. 54, no. 9, pp.
3987-4004, Sep. 2008.

\bibitem{hac-lou-mes-naj-val-2011-rmta} W. Hachem, P. Loubaton, J. Najim, X. Mestre, P. Vallet, 
"Large information plus noise random matrix models and consistent subspace estimation in large 
sensor networks ", Random Matrices, Theory and Applications (RMTA), vol. 1, no. 2 (2012), 
also available on Arxiv (arXiv:1106.5119). 

\bibitem{bilinear-ihp} W. Hachem, P. Loubaton, J. Najim, X. Mestre, P. Vallet, 
"On bilinear forms based on the resolvent of large random matrices", 
Annales Inst. Henri Poincar\'e-Probabilités et Statistiques,  vol.49,  no. 1, pp. 36-63, 
Feb. 2013. 


\bibitem{jiang-stoica-li-2004} Y. Jiang, P. Stoica, J. Li, 
"Array signal processing in the known waveform and steering vector case", 
{\sl IEEE Trans. Signal Process.} vol. 52, no. 1, pp. 23-35, Jan. 2004.

\bibitem{book-kay} S. Kay, "Fundamentals of statistical signal processing,
volume II: detection theory", Prentice Hall, New Jersey 1998.

\bibitem{krichtman-nadler-2009} S. Kritchman, B. Nadler, 
"Non-parametric detection of the number of signals, hypothesis testing and
random matrix theory", {\sl IEEE Trans. Signal Process.}, vol. 57, no. 10, pp.
3930--3941, 2009.

\bibitem{mestre2008modified} X. Mestre, L.A Lagunas, 
"Modified subspace algorithms for DoA estimation with large arrays",
{\sl IEEE Trans. Signal Process.}, vol. 56, no. 2, pp. 598-614, Feb.
2008.

\bibitem{nada-sil10} R.R. Nadakuditi, J.W Silverstein,
"Fundamental limit of sample generalized eigenvalue based detection 
of signals in noise using relatively few signal-bearing and noise-only samples", 
{\sl IEEE J. Sel. Topics Signal Process.}, vol. 4, no. 3, pp. 468
-480, Jun. 2010.
 
\bibitem{rao-edelman-2008} R.R. Nadakuditi, A. Edelman,
"Sample eigenvalue based detection of high-dimensional signals 
in white noise using relatively few samples", 
{\sl IEEE Trans. Signal Process.}, vol. 56, no. 7, pp. 2625-2637, Jul. 2008.

\bibitem{pastur-simple} L.A. Pastur, "A simple approach for the study 
of the global regime of large random matrices", 
Ukrainian Math. J., vol. 57, no. 6, pp. 936-966, Jun. 2005. 

\bibitem{pastur-shcherbina-book} L.A. Pastur, M. Shcherbina, 
"Eigenvalue distribution of large random matrices", 
Mathematical Surveys and Monographs, Providence: American Mathematical Society, 2011. 

\bibitem{book-young-smith} G.A. Young, R.L. Smith, "Essentials of statistical
inference", Cambridge Series in Statistical and Probabilistic Mathematics, 2005. 

\bibitem{p-val-lou-mes-2012} P. Vallet, P. Loubaton, X. Mestre, 
"Improved subspace estimation for multivariate observations of high dimension: 
the deterministic signals case", 
{\sl IEEE Trans. Inf. Theory}, vol. 58, no. 2, pp. 1043-1068, Feb. 2012.

\bibitem{viberg-stoice-ottersten-1997} M. Viberg, P. Stoica, B. Ottersten,
"Maximum likelihood array processing in spatially correlated noise fields using 
parameterized signals", 
{\sl IEEE Trans. Signal Process.} vol. 45, no. 4, pp. 996-1004, Apr. 1997.

\bibitem{zheng-2012} S. Zheng, "Central limit theorems for linear 
spectral statistics of large dimensional F-matrices", {\sl Ann. Inst. Henri Poincar\'e}, 
vol. 48, no. 2, pp. 444-476, 2012.

\bibitem{zhou-et-al-icc-2012} Y. Zhou, E. Serpedin, K. Qarage, O. Dobre, "On the performance
of generalized likelihood ratio test for data aided timing synchronization of MIMO systems", 
Proc. IEEE Int. Conf. Commun. (ICC 2012), Bucharest, pp. 43-46, 2012.

\end{thebibliography}

\end{document}